\documentclass[useAMS,usenatbib]{mn2e}
\usepackage{amssymb,multirow,tabularx}
\usepackage{amsmath}
\usepackage{graphicx}
\usepackage{BibDef}
\usepackage{xcolor}
\usepackage{hyperref}
\hypersetup{
  colorlinks=true,        % false: boxed links; true: coloured links
  linkcolor=blue,         % colour of internal links
  citecolor=blue,         % colour of links to bibliography
}

\usepackage{ulem}
\def\lsim{\mathrel{\rlap{\lower 3pt \hbox{$\sim$}} \raise 2.0pt \hbox{$<$}}}
\def\gsim{\mathrel{\rlap{\lower 3pt \hbox{$\sim$}} \raise 2.0pt \hbox{$>$}}}
\def\msun{\rm {M_{\large \odot}}}

\title[Implications of MBH triplets for PTAs]
{Post-Newtonian evolution of massive black hole triplets in galactic nuclei -- III. A robust lower limit to the nHz stochastic background of gravitational waves.}
\author[Bonetti et al.]{Matteo Bonetti$^{1,2}$, Alberto Sesana$^3$, Enrico Barausse$^{4,5}$ \& Francesco Haardt$^{1,2}$\\
$^1$DiSAT, Universit\`a degli Studi dell'Insubria, Via Valleggio 11, 22100 Como, Italy\\
$^2$INFN, Sezione di Milano-Bicocca, Piazza della Scienza 3, 20126 Milano, Italy\\
$^3$Institute of Gravitational Wave Astronomy and School of Physics and Astronomy, University of
Birmingham, Edgbaston, Birmingham \\ \, B15 2TT, United Kingdom\\
$^4$CNRS, UMR 7095, Institut d'Astrophysique de Paris, 98 bis Bd Arago, 75014 Paris, France\\
$^5$Sorbonne Universit\'es, UPMC Univesit\'e Paris 6, UMR 7095, Institut d'Astrophysique de Paris, 98 bis Bd Arago, 75014 Paris, France
}

\begin{document}

\date{~}

\pagerange{\pageref{firstpage}--\pageref{lastpage}} \pubyear{2017}

\maketitle

\label{firstpage}

%%%%%%%%%%%%%%%%%%%%%%%%%%%%%%%%%%%%%%%%%%%%%%%%%%%%%%%%%%%%%%%%%%%%%%%%%%%%%%%%%%%%%
\begin{abstract} 
Inspiraling massive black-hole binaries (MBHBs) 
forming in the aftermath of galaxy mergers are expected to be the loudest gravitational-wave (GW) 
sources  relevant for pulsar-timing arrays (PTAs) at nHz frequencies. The incoherent overlap of signals 
from a cosmic population of MBHBs gives rise to a stochastic GW background (GWB) with characteristic strain 
around $h_c\sim10^{-15}$ at a reference frequency of 1 yr$^{-1}$, although uncertainties around this value are large. 
Current PTAs are piercing into the GW amplitude range predicted by MBHB-population models, but no 
detection has been reported so far. To assess the future success prospects of PTA experiments, it is therefore important 
to estimate the minimum GWB level consistent with our current 
understanding of the formation and evolution of galaxies and massive black holes (MBHs). To this purpose, we couple a semianalytic model 
of galaxy evolution and an extensive study of the statistical outcome of triple MBH interactions. We show that even in the 
most pessimistic scenario where all MBHBs stall before entering the GW-dominated regime, triple interactions resulting from subsequent 
galaxy mergers inevitably drive a considerable fraction of the MBHB population to coalescence. At frequencies relevant for PTA, the 
resulting GWB is only a factor of 2-to-3 suppressed compared to a fiducial model where binaries are allowed to merge over Gyr timescales. 
Coupled with current estimates of the expected GWB amplitude range, our findings suggest that the minimum GWB from cosmic MBHBs is unlikely to 
be lower than $h_c\sim10^{-16}$ (at $f = 1$ yr$^{-1}$), well within the expected sensitivity of projected PTAs based on future observations with FAST, MeerKAT and SKA.
\end{abstract}
%%%%%%%%%%%%%%%%%%%%%%%%%%%%%%%%%%%%%%%%%%%%%%%%%%%%%%%%%%%%%%%%%%%%%%%%%%%%%%%%%%%%%


%%%%%%%%%%%%%%%%%%%%%%%%%%%%%%%%%%%%%%%%%%%%%%%%%%%%%%%%%%%%%%%%%%%%%%%%%%%%%%%%%%%%%
%%%%%%%%%%%%%%%%%%%%%%%%%%%%%%%%%%%%%%%%%%%%%%%%%%%%%%%%%%%%%%%%%%%%%%%%%%%%%%%%%%%%%
\section{Introduction}
\label{sec:Intro}
Massive black holes (MBHs) are among the primary building blocks of galaxies. Since the early nineties it has become clear that most (if not all) 
massive galaxies host a MBH at their centre \citep{Kormendy1995} whose mass correlates with the properties of the galactic host \citep{Magorrian1998,Ferrarese2000,Gebhardt2000}, 
pointing towards a co-evolution of MBHs and their host galaxies. In the standard hierarchical framework of structure formation, galaxies form in a bottom-up fashion, whereby the massive galaxies that we see today
build up at the intersection of dark matter filaments along which galaxies and  cold gas can stream inwards without getting shock-heated
\citep{Dekel2009b}.
At those intersections, galaxies experience a sequence of mergers and accretion
events, that contribute to their final mass. If MBHs are ubiquitous in galaxy centres, then MBH binaries (MBHBs) naturally form following galaxy mergers \citep{Begelman1980}. If MBHs can efficiently pair 
into sub-parsec binaries, then gravitational wave (GW) emission inevitably takes over \citep{Peters1963}, leading to final coalescence. In this scenario, MBHs grow through a series of mergers and accretion 
events, and the interplay of accretion and feedback is thought to be responsible of the observed MBH-galaxy relations \citep{Kauffmann2000,Volonteri2003}.

MBHBs are the loudest GW sources in the Universe, and a typical billion-solar-mass system inspiralling at centi-parsec scales emits GWs at nHz frequencies \citep{Sesana_Vecchio2008}, currently probed by pulsar-timing 
arrays \citep[PTAs,][]{Foster1990}. In fact, milli-second pulsars (MSPs) are the most stable macroscopic clocks known in Nature \citep{Taylor1992}, and a GW passing through their line of 
sight leaves a characteristic imprint on the time of arrivals (TOAs) of the pulses on Earth \citep{Sazhin1978}. By working within the hierarchical structure-formation paradigm, it is possible to 
construct the expected distribution of MBHBs populating the Universe, as a function of mass, redshift and orbital frequency. The GW signals from individual binaries will add up incoherently to form a stochastic 
GW background \citep[GWB,][]{Rajagopal1995,Jaffe2003,Wyithe2003,Sesana2004}, which in its simplest form (i.e. assuming a population of circular, GW driven MBHBs) has a single power-law spectrum with characteristic
strain parametrised as $h_c=A[f/(1 \ {\rm yr}^{-1})]^{-2/3}$ \citep{Phinney2001}, where $f$ is the observed frequency. Note that essentially all results in the PTA literature are quoted as strain at  $f = 1$ yr$^{-1}$, we therefore quote our bulk results in terms of the normalisation $A$. 

The amplitude of the signal has been computed by several authors in the past 
two decades using several approaches, including: analytic and semianalytic models for structure formation and the MBHB merger rate \citep{Wyithe2003,Sesana_Vecchio2008,McWilliams2014},
empirical galaxy merger rate estimates from observations \citep{Rajagopal1995,Jaffe2003,Sesana2013,Ravi2015,Sesana2016} and 
large cosmological simulation \citep{Sesana2009,Ravi2012,Kulier2015,Kelley2017}. The typical value of $A$ inferred 
from these theoretical models is of the order of $\sim 10^{-15}$. This is particularly interesting, because $10^{-15}$ is the sensitivity 
that is achievable by timing about ten MSPs with a precision of about $100$ ns over several years \citep{Jenet2006}, 
which is now within reach of the leading PTA experiments. In fact the European PTA \citep[EPTA,][]{Desvignes2016}, NANOGrav \citep{NANOGrav2015}, the Parkes PTA \citep[PPTA,][]{Reardon2016} 
and the International PTA \citep[IPTA,][]{Hobbs2010} have recently placed upper limits on the amplitude of a stochastic GWB at a level of 
$A=3\times 10^{-15}, 1.5\times 10^{-15}, 1\times 10^{-15}, 1.7\times 10^{-15}$ respectively \citep{Lentati2015,Arzoumanian2016,Shannon2015,Verbiest2016}.

PTAs are becoming a world-wide effort that requires the investment of significant human resources and radio-telescope time, including upcoming facilities like FAST, MeerKAT and, eventually, 
the SKA \citep[see][for future prospects]{Janssen2015}. It is therefore of paramount importance to carry out a comprehensive exploration of the predicted GW signal in order to critically assess the 
chances of success of this endeavour. The aforementioned theoretical models considered a wide range of scenarios, differing in several physical ingredients, including: i) a range of 
galaxy-evolution models and underlying galaxy-merger rates; ii) several MBH-galaxy scaling relations; iii) the role of gas and stellar dynamics in driving the MBHB at low frequencies; iv) eccentricity. 
In general, however, there is a common feature to all those models: MBHBs can efficiently reach the centi-parsec separations relevant for PTA observations \citep[notably,][explored a range of efficiencies in the context of stellar driven MBHBs]{Kelley2017}. Although this is a reasonable assumption, MBHs have to undergo a long journey before getting to those
small separations. 
In fact, following galaxy mergers, 
dynamical friction is only efficient in driving the two MBHs  to form a Keplerian binary \citep{Begelman1980}. For billion-solar-mass systems, this happens at separations of the order of tens of
parsecs, where GW emission is still inefficient. Hardening driven by interactions with the dense stellar environment of galactic nuclei \citep[e.g.,][]{Khan2012,Sesana2015,Vasiliev2015}
or by a putative circumbinary disc \citep[e.g.,][]{MacFadyen2008,Cuadra2009,Nixon2011} can bridge the gap, taking the MBHBs down to centi-pc separation. However, the efficiency of these processes 
critically depends
on a number of physical conditions, such as the stellar density in the nuclei of very massive galaxies, the efficiency of relaxation processes 
in bringing stars on almost radial orbits that intersect the MBHB, or simply
the mere availability of cold gas to form a sizeable circumbinary disc.
This is sometimes referred to in the literature as 
the ``final-parsec problem''~\citep[see][for an overview of general issues related to bound MBHB evolution]{Dotti2012}.

It is therefore of great interest to ask the following question: what if all hardening mechanisms fail and MBHBs simply stall? Would that be the tombstone of PTA experiments? This scenario has recently been explored
by \citet{Dvorkin2017} (hereinafter DB17); they argued that even if all binaries stall, there would be a (much reduced) leftover GW signal in the PTA band. This is because the typical MBHB stalling 
radius is a function of the binary mass and mass ratio, and even though billion-solar-mass systems stall well outside the PTA band, lighter MBHB and low mass-ratio ones can still emit some GWs at nHz frequencies. 
The signal they predict is, however,
at a level of $A\lesssim 10^{-16}$. 
A crucial ingredient that was missing in the DB17 modelling, however, is the formation of multiple MBH systems. In fact, massive galaxies typically experience multiple mergers along their formation history 
\citep[see, e.g.,][]{Rodriguez-Gomez2015}. If a MBHB stalls, the subsequent merger event will bring a third MBH, thus forming a MBH triplet 
\citep[][Paper I hereinafter]{Volonteri2003,Iwasawa2006,Hoffman2007,Kulkarni2012,Bonetti2016}.

In this paper we explore the effect of MBH triple and quadruple interaction on the GW signal
generated by the cosmic MBH population in the PTA band\footnote{While completing this draft, we became aware of a similar
independent investigation by Ryu and collaborators \citep{Ryu2017}. Albeit employing different frameworks, the two studies reach similar conclusions}. 
We model the statistics of MBH binaries, triplets and quadruplets by coupling the large suite of simulations described in \citet{Bonetti2017b} (Paper II hereinafter) 
to the semianalytic model of galaxy formation of \citet{Barausse2012}. We consider two conceptually different situations: a fiducial model in which MBHBs merge on 
timescales of millions-to-billions year (depending on galactic properties) after their host galaxies merge, and an extreme model in which all MBHBs stall at about their hardening radius (see Paper II for details), and mergers are prompted only by
multiple MBH interactions. Within the context of the assumed galaxy-formation model, the latter is the most pessimistic scenario from the point of view of the expected GW 
signal in the PTA band. The difference between the two scenarios is indicative of the typical maximum suppression of the GW signal, for a given galaxy formation model, due 
to MBHB stalling. We show that multiple MBH interactions prompt the coalescence of a number of MBHBs 
which results, in the relevant 
PTA frequency band, in a GWB only a factor 2-to-3 reduced with respect to the efficient binary merger case.
This is to be compared to the suppression factor of 10 or larger found by DB17, which neglected multiple MBH interactions.
Our results imply that even a combination of a particularly 
unfavourable MBH-galaxy scaling relations \citep{Sesana2016} and MBHB stalling still results in a GWB at the $A\approx 10^{-16}$ level, well within the capabilities of a realistic SKA-era PTA. 

The paper is organised as follows. In Section \ref{sec:semianalytic} we describe the MBH evolution model and the implementation of triple and quadruple interactions. 
Section \ref{sec:GW} summarises the method for the GWB computation. Our main results are presented in Section \ref{sec:results}, and a few important caveats are discussed in
Section \ref{sec:discussion}. We conclude with some final remarks and directions for future investigation in Section \ref{sec:conclusions}. We assume a concordance $\Lambda$--CDM 
universe with $\Omega_M=0.3$, $\Omega_\Lambda=0.7$, and $H_0=70$ km/(s Mpc). Unless otherwise specified, we use geometric units where $G=c=1$.

%%%%%%%%%%%%%%%%%%%%%%%%%%%%%%%%%%%%%%%%%%%%%%%%%%%%%%%%%%%%%%%%%%%%%%%%%%%%%%%%%%%%%
%%%%%%%%%%%%%%%%%%%%%%%%%%%%%%%%%%%%%%%%%%%%%%%%%%%%%%%%%%%%%%%%%%%%%%%%%%%%%%%%%%%%%
\section{Semianalytic model of galaxy and massive black hole evolution}
\label{sec:semianalytic}

%%%%%%%%%%%%%%%%%%%%%%%%%%%%%%%%%%%%%%%%%%%%%%%%%%%%%%%%%%%%%%%%%%%%%%%%%%%%%%%%%%%%%
\subsection{General description of the model}
\label{sec:description}

We simulate the co-evolution of MBHs and their host galaxies by the semianalytic galaxy-formation model of~\citet{Barausse2012},
with later incremental improvements described in~\citet{Sesana2014,Antonini2015,Antonini_Barausse2015}. We refer to those references for a detailed description of the model. 
Here, we limit ourselves to summarising its most salient features. The model's calibration is described in \citet{Barausse2017}, and reproduces the conservative MBH scaling relations of ~\citet{Shankar2016},
which are known to produce a low level of the stochastic GW signal for PTAs~\citep{Sesana2016}.

The evolution of Dark-Matter halos is modelled via merger trees produced with an extended Press-Schechter formalism, suitably
modified so as to reproduce the results of N-body simulations~\citep{Press1974,Parkinson2008}. The baryonic components of galaxies are then evolved along
the branches of these merger trees, while the nodes of the trees
correspond to the moment when two halos touch, thus initiating the processes leading to halo, galaxy and eventually black-hole coalescence. 

In more detail, galaxies form from either the cooling of an unprocessed ``hot'' gas component shock-heated to
the halo's virial temperature, or (especially in low-mass systems and at high redshift) from accretion flows of colder gas \citep{Dekel2006,Cattaneo2006,Dekel2009}. 
When the gas has reached the halo's centre as a result of
either of these channels, it forms a disc, simply by conservation of angular momentum, and eventually starts forming stars. Galactic spheroids form instead when the
gaseous and galactic discs are destroyed either by bar instabilities or by major galactic mergers. Both these processes are also assumed to
drive cold gas to the galactic centre, thus enhancing star formation. During star formation episodes 
(in both discs and spheroids) we also account for the feedback from supernova explosions on the surrounding gas.

MBHs are formed from high-redshift seeds, with several possible plausible choices for their 
initial mass function and  halo occupation fraction. In this work, we consider a ``light-seed'' (LS) scenario,
where seeds of a few hundred $M_\odot$ are provided by the remnants of popIII stars forming in low-metallicity high-redshift galaxies~\citep{Madau2001}, and a ``heavy-seed'' (HS) scenario 
where larger 
($\sim 10^5 M_\odot$) seeds form from the collapse (e.g. due to bar instabilities) of protogalactic discs. More precisely, in the LS scenario 
we populate with seeds only the most massive halos (i.e. those collapsing from the $3.5\sigma$ peaks of the primordial density field)
at redshift $15<z<20$. The mass of each seed is assumed to be $\sim 2/3$ of the initial popIII star's mass (to account for mass losses during the collapse of the star). The mass
of the star is drawn randomly from a log-normal distribution
centred  at  300 $M_\odot$ with rms  of  0.2  dex
and  an  exclusion  region  between  140  and  260 $M_\odot$, since stars in this range explode as pair-instability supernovae without forming black holes~\citep{Heger2002}.
For the HS scenario, we follow \citet{Volonteri2008}, which
models the formation of seeds from disc bar instabilities at redshift $15<z<20$. The model has just one free parameter, i.e. the critical Toomre parameter
$Q_c$ at which the instability sets in. The most likely values for $Q_c$ range between 2 and 3. Here, we adopt $Q_c=2.5$.

The black-hole seeds then grow via accretion and mergers. For the former channel, we assume that a gas reservoir forms in the nuclear region of each galaxy
as a result of cold gas being funnelled to the centre during major galactic mergers or bar instabilities of the gaseous galactic discs. Because both kinds of events are also thought
to trigger spheroid formation, we follow~\citet{Granato2004,Lapi2014} and assume that the feeding of this nuclear reservoir is 
linearly correlated with the star formation rate in the spheroid component.
The MBH then accretes from this reservoir on the viscous timescale, but we cap this accretion rate at the Eddington rate in the HS scenario, and at twice the Eddington rate
in the LS one.~\footnote{This choice is made because a certain degree of super-Eddington accretion is needed in 
an LS scenario, if one wants to reproduce the high-redshift AGN
luminosity function, c.f.~\citet{Madau2014}.} 
As a result, MBHs undergo periods of quiescent activity interrupted by
quasar/AGN phases. The feedback of the MBH on 
the surrounding gas is taken into account in both phases (radio-mode and quasar feedback).
The nuclear reservoir is also assumed to form stars, which give rise to a nuclear star cluster~\citep{Antonini2015,Antonini_Barausse2015}. We also assume that nuclear star clusters 
form and  grown via a 
second channel, namely the migration of globular clusters to the nuclear region induced by dynamical friction~\citep{Antonini2015,Antonini_Barausse2015}.

After two halos start coalescing (at the nodes of the merger tree), the smaller halo (the ``satellite'') initially retains its identity within 
the bigger one (the ``host''), slowly falling towards the centre
driven by dynamical friction. We account for this phase by using the expression for the dynamical-friction time from \citet{Boylan-Kolchin2008}, 
which is calibrated against numerical simulations and
 accounts for the effect of both Dark Matter and baryons. We also model the mass loss incurred by the satellite halo and its galaxy due to tidal 
 stripping and evaporation~\citep{Taffoni2003}.
 When the satellite subhalo and its galaxy merge with the host, the MBHs contained in the two galaxies may still be very far apart (at $\sim$ kpc distances). Nevertheless, at
 least when the satellite and the host have mass ratios $\gtrsim 0.1$ (the most relevant case for our results, c.f. section~\ref{sec:discussion}), 
 dynamical friction against the gas and stars 
 of the newly formed galaxy quickly drives the MBHs toward the centre~\citep[DB17,]{Dosopoulou2017}. This process is particularly efficient because at least in its initial stages, 
 the MBHs are expected to be still surrounded by a stellar core from their host galaxy, which results in a shorter dynamical-friction timescale. Therefore, for the purpose of this work,
 we can safely neglect this phase and assume that whenever the host and satellite galaxies coalesce, 
 the MBHs are efficiently driven down to a separation comparable to the primary MBH influence radius $r_i\approx 2Gm_1/\sigma^2$
($m_1$ being the primary black hole's mass and $\sigma$ the velocity dispersion of the host's spheroid).

 We then account for the subsequent evolution of the MBHs with the simple prescriptions outlined in~\citep{Antonini2015,Antonini_Barausse2015}. In particular, we discriminate between MBHB mergers driven by gas or by stars by computing the mass of the gas in the central pc-sized nucleus of the galaxy. If the gas mass exceeds the mass of the  MBHB, we assume that the merger occurs in a gas-rich environment and we
 assume that the MBH binary
 is driven to sub-pc separation (where GW emission is sufficiently efficient to trigger the merger) by planetary-like migration
 within the nuclear disc. We consider that migration proceeds on the viscous timescale (evaluated at the influence radius of the binary), 
 i.e. $\sim 10^7-10^8$ yr. In the opposite case, i.e., when the gas mass is negligible (a situation more relevant for PTAs), 
 the MBH binary is instead driven to sub-pc separations by stellar hardening, i.e., by
 three-body interactions with stars. We model this phase by assuming that its duration is given by the hardening timescale evaluated at the influence radius of the MBH 
 binary~\citep{Sesana2015}, or to the timescale of the hardening 
 from the nuclear star cluster, whichever the shorter. In practice, these hardening timescales are typically of a few Gyr.
 We have checked that our results do not depend sensitively on the employed prescription to discriminate gas-rich from
 gas-poor mergers. In practice, one finds that the vast majority of MBHB mergers is
 gas-poor in the PTA band (see table~\ref{tab2}).

 We account for the possibility that while a MBH binary is still evolving under gas-driven migration or hardening, another galaxy merger may take place.
This
would bring a third MBHs down to pc-separations, which may in turn trigger the coalescence of the inner binary via Kozai-Lidov resonances~\citep{Kozai1962,Lidov1962} 
or chaotic three-body interactions (Paper I). 
We model these processes by using the results of the three-body Post-Newtonian (PN) code of Paper I, 
as outlined in the next section.
Clearly, the importance of these MBH triple systems in triggering MBH coalescences depends sensitively on the ``delays'' between the time halos start merging and the time MBHs eventually 
coalesce. In the
following, we consider both a case where these delays are implemented as described above ({\it Model-delayed} hereinafter) and a case where
the delays are artificially set to values larger than the age of the Universe (i.e., MBHBs stall at about their hardening radius, therefore mergers never take place unless triggered by three-body MBH interactions, {\it Model-stalled} hereinafter). 
Coupled with the different MBH seeding prescription, we therefore have four distinct models: {\it Model-delayed-LS},  {\it Model-stalled-LS}, {\it Model-delayed-HS},  {\it Model-stalled-HS}. 
We find that low and high mass seeds ({\it LS} vs {\it HS}) yield very similar results in the MBH mass range relevant to PTAs \citep[but not at the low masses relevant to LISA,][as we will explore in a forthcoming paper]{Klein2016}. Unless otherwise stated, we will therefore always show results for the {\it LS} models.    

%%%%%%%%%%%%%%%%%%%%%%%%%%%%%%%%%%%%%%%%%%%%%%%%%%%%%%%%%%%%%%%%%%%%%%%%%%%%%%%%%%%%%
\subsection{Treatment of triple and quadruple MBH systems}
\label{sec:triplets}

%%%%%%%%%%%%%%%%%%%
\begin{figure}
\includegraphics[scale=0.35]{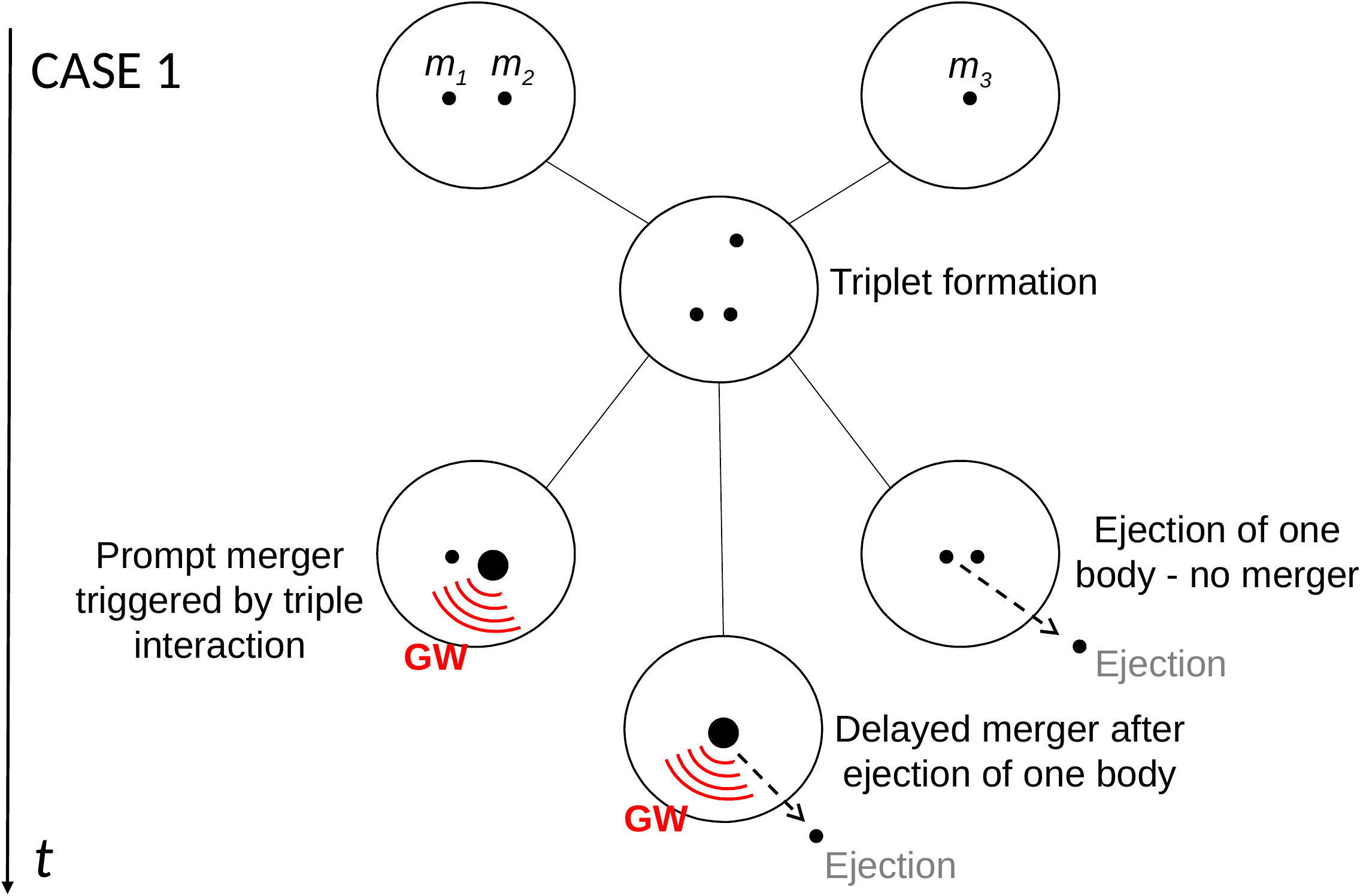}
\caption{Cartoon representation of how triple MBH interactions are treated in the semianalytic model described in  Section \ref{sec:semianalytic}.}
\label{fig:triplet1}
\end{figure}
%%%%%%%%%%%%%%%%%%%
\begin{figure}
\includegraphics[scale=0.35]{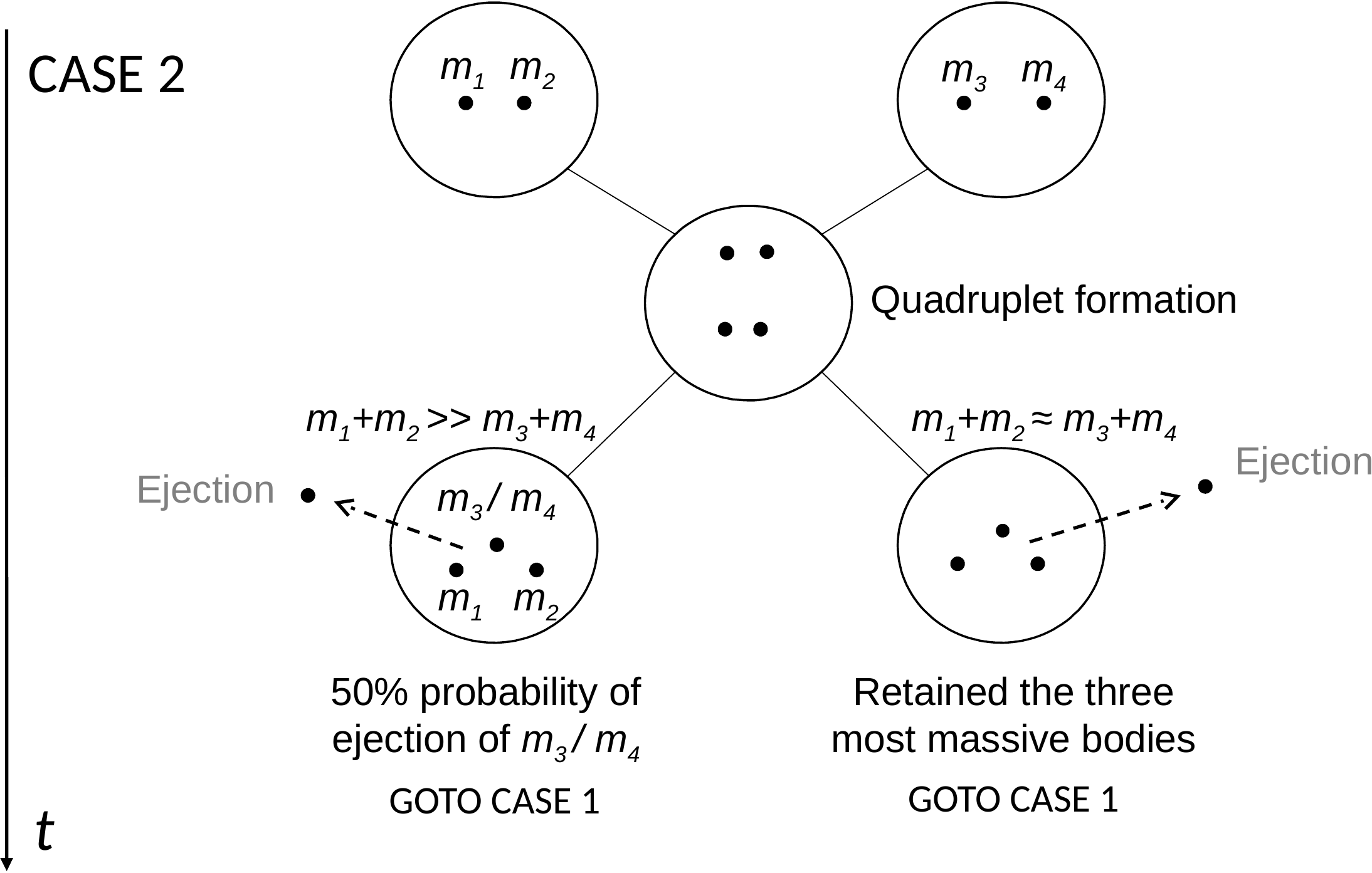}
\caption{Same as figure~\ref{fig:triplet1}, but for quadruple interactions.}
\label{fig:triplet2}
\end{figure}
%%%%%%%%%%%%%%%%%%%

The backbone for a consistent treatment of multiple (i.e. triple and quadruple) MBH interactions in the semianalytic model described above is 
the large suite of numerical simulations described in Paper II. Full details can be found in there, and we only describe here the key features and the
implementation within the semianalytic galaxy evolution model. We have collected the outcome of triple MBH interactions on a grid of primary MBH mass,
$m_1$, sampled in the range $[10^5\msun,10^{10}\msun]$, and inner and outer binary mass ratio $q_{\rm in}\in[0.03,1]$and $q_{\rm out}\in[0.03,10]$\footnote{While we define $q_{\rm in}\leq 1$ 
by construction, the intruder might be more massive than the pre-existent MBHB (even though this occurs in a minority of cases). Note that for $q_{\rm out}>1$ we performed a restricted 
set of simulations for $m_1=10^9\msun$ only. Given that none of the results for $q_{\rm out}<1$ has a strong dependence on the mass scale, we extrapolate the $q_{\rm out}>1$ results obtained 
for $m_1=10^9$ to all masses.}. For each point in the 3D grid ($m_1$, $q_{\rm in}$, $q_{\rm out}$), we simulate several systems with different inner and outer orbit eccentricities, and different relative 
inclinations, and we use the results to compute merger fractions, merger-time distributions and MBHB eccentricity distributions just before merger (more precisely, we record eccentricities at separations of $100R_{\rm G}$, where
$R_G=GM/c^2$ is the gravitational radius associated with the merging MBHB total mass $M=m_1+m_2$). In particular, we isolate three distinct outcomes (see figure \ref{fig:triplet1}) and their associated
occurrence probabilities:

%%%%%%%%%%%%%%%%%%%
\begin{enumerate}
\item A prompt coalescence triggered by a triple interaction. The coalescence can involve any one pair of MBHs in the triplet. We identify here with body 1 and 2 the two MBHs of the inner binary ($m_1>m_2$ by definition) 
and with body 3 the intruder. For each grid point in our simulation suite, we record three numbers $a$, $b$ and $c$ identifying the fractions of simulations in which bodies 1-2, 1-3 and 2-3 merge, respectively.
\item An ejection of one of the MBHs (the lighter, in the overwhelming majority of the cases) and a delayed merger of the remaining binary (shrunk or made more eccentric by the 3-body interaction) under the effect of GW emission. 
This occurs in a fraction $d$ of the realizations.
\item An ejection and a left-over binary unable to merge alone within the Hubble time. Such a binary is then retained in the semianalytic model as it can potentially undergo new multiple MBH interactions following
later galaxy mergers, or coalesce under the effect of gas-driven migration or stellar hardening (in {\it Model-delayed} only). 
\end{enumerate}
%%%%%%%%%%%%%%%%
The relative occurrence of the different outcomes depend on the chosen point in the 3-D grid ($m_1$, $q_{\rm in}$, $q_{\rm out}$). We also
stress that our treatment of triple interactions is conservative, because the ejected MBH may fall back to the galactic nucleus after the left-over binary has merged, thus
potentially providing an additional MBH merger in a minority ($\sim 10-20 \%$) of cases~\citep{Hoffman2007}.

In any given triple interaction produced by the semianalytic model, the probability of a given outcome is obtained by using a trilinear interpolation between the surveyed grid 
points, to estimate the fractions $a, b, c, d$ for that specific system. A random number $P$ between $0$ and $1$ is then drawn and, according to its value, 
one of the following choices is selected: 

%%%%%%%%%%%%%%%%%%%%%%%%%%%%%%
\begin{itemize}
\item If $P < a+b+c \rightarrow$ prompt merger:
	\begin{itemize}
	\item[-] If $P \leq a \rightarrow$ merger of bodies 1-2.
	\item[-] If $a < P \leq a+b \rightarrow$ merger of bodies 1-3.
	\item[-] If $a+b < P \leq a+b+c \rightarrow$ merger of bodies 2-3.
	\end{itemize}
\item If $a+b+c < P \leq a+b+c+d \rightarrow$ delayed merger (of the two most massive bodies).
\item If $a+b+c+d < P  \rightarrow$ no merger (left-over binary formed by the two most massive bodies).
\end{itemize}
%%%%%%%%%%%%%%%%%%%%%%%%%%%%%%

In the case of a merger, its timescale is obtained by sampling the distribution of merger timescales for the two cases of prompt and delayed mergers respectively (see Paper II). 
At this stage, we do not follow self consistently the eccentricity evolution of individual MBHB systems in the semianalytic model (see next section). 
We also note that the parameters of a given triplet can lie outside the grid sampled in Paper II. In particular we can have $q_{\rm in}<0.03$ and/or $q_{\rm out}<0.03$ or $q_{\rm out}>10$. 
In this case we simply apply the fractions and timescale distributions of the closest grid point. Although this is certainly a crude approximation, 
the GW signal is much weaker for low mass-ratio binaries, and our treatment of those systems does not change our results significantly (see section \ref{sec:discussion}).  

Besides the formation of triplets, quadruple interactions (caused by the merger of two galaxies, both hosting a binary) are a natural occurrence, especially in {\it Model-stalled}. 
In absence of a library of simulations of quadruple interactions, we reduce the problem to the triplet case, as shown in figure~\ref{fig:triplet2}. 
If one of the two binaries is much lighter than the other (we arbitrarily choose a threshold mass ratio of 0.1), 
we expel one of its two members with $50\%$ probability, irrespectively of the binary's mass ratio, 
retaining only one intruder and reducing the problem to the triplet case. This assumption is made in analogy to the problem of a stellar binary interacting with a much more massive 
object \citep[usually a MBH or an intermediate MBH,][]{Bromley2006}. If the two binaries have comparable total mass (mass ratio larger than 0.1), 
we retain the three more massive bodies, again reducing the problem to the triplet case. We stress that this assumption is conservative, mostly because it neglects the possibility of multiple mergers. 
If the four MBHs form a hierarchical system of two binaries, for example, Kozai-Lidov oscillations might induce mergers of both binaries. 

Each of our semianalytic models thus produces a catalogue of MBH mergers containing the masses of the two merging MBHs and the merger's redshift. 
If the merger involved a standard MBHB, we flag the event either as `star' or `gas' depending on whether the binary evolved in a stellar (i.e. gas-poor) or gaseous (i.e. gas-rich) environment. 
If the merger was instead triggered by a multiple (triple or quadruple) interaction, we also record $q_{\rm in}$ and $q_{\rm out}$ of the progenitor triplet, 
and we flag the system as `Tr' if the merger was promptly triggered during the triple interaction, or as `Tr-ej' if the merger was driven by GW emission after the 
ejection of one MBH during the triple interaction.  In the `Tr' case, we also record whether the progenitor system was originally a triple or a quadruple, 
to assess the relative importance of the two populations. Those catalogues are used to construct differential distributions of merging binaries, which we use to  compute the GW signal, as detailed in the next section.

%%%%%%%%%%%%%%%%%%%%%%%%%%%%%%%%%%%%%%%%%%%%%%%%%%%%%%%%%%%%%%%%%%%%%%%%%%%%%%%%%%%%%
\section{Computation of the gravitational-wave signal}
\label{sec:GW}
We adopt two different techniques for the computation of the GWB in the case of `regular' mergers and mergers induced by triple interactions. 
For regular MBHBs, we assume circular orbits for simplicity. Therefore, the characteristic strain can be expressed as:

%%%%%%%%%%%%%%%%%%%%%%%%%%%%%%%%%
\begin{equation}
h_c^2(f) =\frac{4}{\pi f^2}\int dz \int dm_{1} \int
dq \, \frac{d^3n}{dzdm_{1}dq}
{1\over{(1+z)}}~{{dE_{\rm gw}} \over {d\ln{f_r}}}\,.
\label{hcdE}
\end{equation}
%%%%%%%%%%%%%%%%%%%%%%%%%%%%%%%%%
%
where $d^3n/(dzdm_1dq)$ is the differential number density of MBHB merger per unit redshift, primary MBH mass and binary mass ratio constructed from the output catalogues of the semianalytic model, and the differential energy spectrum $dE_{\rm gw}/d\ln{f_r}$ can be broken down as

%%%%%%%%%%%%%%%%%%%%%%%%%%%%%%%%%
\begin{equation}
\frac{dE_{\rm gw}}{d\ln{f_r}}=\frac{dE_{\rm gw}}{dt_r}\frac{dt_r}{df_r}f_r.
\label{dedlnf}
\end{equation}
%%%%%%%%%%%%%%%%%%%%%%%%%%%%%%%%%
%
Note that time and frequency are evaluated in the source rest-frame, so that compared to the time and frequency at the observer, we have $t_r=t/(1+z)$ and $f_r=f(1+z)$. The first term on the right-hand side 
of equation (\ref{dedlnf}) is the GW luminosity, given by

%%%%%%%%%%%%%%%%%%%%%%%
\begin{equation}
\frac{dE_{\rm gw}}{d t_r}=\frac{32}{5}{\cal M}^{10/3}(\pi f_r)^{10/3},
\label{eq:dEdt}
\end{equation}
%%%%%%%%%%%%%%%%%%%%%%
%
where ${\cal M}=(m_1 m_2)^{3/5}/(m_1+m_2)^{1/5}$ is the binary's chirp mass. The term $dt_r/df_r$ represents the time a given binary spends emitting at a specific frequency $f_r$. 
The main contributors to the GWB are moderately heavy binaries merging in massive galaxies at relatively low redshift. Since most of those mergers are gas poor (cf table \ref{tab2}), 
we assume that MBHBs evolve exclusively because of three-body interactions against the stellar environment and emission of GW, and we can write

%%%%%%%%%%%%%%%%%%%%%%
\begin{equation}
\frac{df_r}{d{t_r}}=\frac{df_r}{dt_r}\Big{|}_{3b}+\frac{df_r}{dt_r}\Big{|}_{\rm gw}=\mathcal{A}f^{1/3}+\mathcal{B}f^{11/3},
\label{eq:f_evol}
\end{equation}
where \citep{Chen2017}
\begin{align}
  \mathcal{A} & = \frac{3}{2 \pi^{2/3}} \frac{H\rho_i}{\sigma} M^{1/3},
  \label{eq:fstar}
\\
  \mathcal{B} & = \frac{96}{5} (\pi)^{8/3} \mathcal{M}^{5/3}.
  \label{eq:fgw}
\end{align}
%%%%%%%%%%%%%%%%%%%%%%
%
Here, $M=m_1+m_2$ is the MBHB total mass, $\sigma$ is the velocity dispersion of the host galaxy, $\rho_i$ is the stellar density at the binary's influence radius 
(both of which are evaluated from the output of the semianalytic galaxy-formation model),
 and $H\approx 15$ is a numerical factor calibrated against dedicated three-body scattering experiments \citep{Quinlan1996,Sesana2006}. The binary evolution is dominated by three-body scattering in the 
 early phase, whereas GW emission takes over at higher frequencies. The transition frequency $f_t$ can be computed by equating the two contributions to $df/dt$, and for typical PTA sources lies in the nHz range. 

Triple interactions can result in extremely eccentric binaries, making the analytic computation of the GWB somewhat time consuming \citep[see however,][]{Taylor2017}. 
Moreover, the eccentricity evolution can be highly chaotic (Paper I), and thus the construction of a simple analytic $dE/df$ is not possible. We note, 
however, that in the PTA frequency range MBHBs show a rather regular behaviour, following $f-e$ tracks dictated by GW backreaction and hardly affected by the third body  (cf figure 9 in Paper II). 
We can therefore consider, to first approximation, a population of eccentric MBHBs evolving solely because of GW emission. \citet{Chen2017} showed that the stochastic GWB for an arbitrary 
population of GW-driven eccentric binaries can be simply obtained by evaluating the single spectrum of a fiducial system, and rescaling it appropriately to match the parameter of the considered sources. 
The total GWB can be written as:

%%%%%%%%%%%%%%%%%%%%%%%
\begin{equation}
\begin{split}
  h_c^2(f) = & \int dz \int dm_1 \int dq \int de \frac{d^4n}{dzdm_1dqde} h_{c,{\rm fit}}^2\Big(f\frac{f_{p,0}}{f_{p}}\Big) \\ & \Big(\frac{f_{p}}{f_{p,0}}\Big)^{-4/3} \Big(\frac{\mathcal{M}}{\mathcal{M}_0}\Big)^{5/3} \Big(\frac{1+z}{1+z_0}\Big)^{-1/3},
  \label{eq:hcanalytic}
\end{split}
\end{equation}
%%%%%%%%%%%%%%%%%%%%%%%
%
where $d^4n/(dzdm_1dqde)$ is now the differential number density of MBHB mergers per unit redshift, primary mass, mass ratio and eccentricity. This quantity is constructed from the catalogue produced by the 
semianalytic model. Since we do not follow the eccentricity evolution of MBHBs self-consistently, for each event we draw a value of $e$ from the distribution corresponding to the appropriate parent triplet properties, 
interpolating the distributions obtained at the grid points of our suite of numerical integrations (cf section \ref{sec:triplets};  as mentioned earlier, eccentricity distributions are recorded at a 
reference binary separation of $100R_{\rm G}$). $h_{c,{\rm fit}}$ is a reference spectrum for a binary with parameters $(\mathcal{M}_0, z_0, f_0, e(f_0))$, which is adapted to arbitrary MBHB parameters 
via the scaling factors reported in parenthesis. The factor $f_{p,0}/f_{p}$ is the ratio of the peak frequencies of the two binary spectra. An eccentric binary, in fact, has a peak in the emission spectrum 
that is uniquely determined by specifying $e(f)$ at a given frequency $f$ \citep[details given in][]{Chen2017}. Therefore, if we know $e$ at $100R_G$, we can compute $f_p$ and rescale the fiducial 
spectrum accordingly. We also recall that triple interactions can result in either a `Tr' or `Tr-ej' merger, as described in the previous section.
In general, `Tr' and `Tr-ej' events have very different eccentricity probability distributions, and we therefore distinguish between the two cases and sample from the respective distributions. 

Although qualitatively different, the two GWB computations of equations (\ref{hcdE}) and (\ref{eq:hcanalytic}) are perfectly consistent with each other. 
We have checked that by artificially setting arbitrarily small $e$ in the triplet population, the GWB obtained via equation (\ref{eq:hcanalytic}) coincides with that obtained via equation (\ref{hcdE}), 
assuming purely GW-driven circular binaries.

Summarising, to practically evaluate the GWB in the two models, we proceed as follow. We flag the origin of each MBHB as merging because of:

%%%%%%%%%%%%%%%%%%%%
\begin{enumerate}
\item Standard dynamical processes (flag 'star').
\item Dynamical processes during a triple interaction (flag 'Tr').
\item GW emission after the ejection of the lighter MBH involved in the triple interaction (flag 'Tr-ej').
\end{enumerate}
%%%%%%%%%%%%%%%%%%%%
%
For each subset of systems we construct the relevant differential number density $d^3n/(dzdm_1dq)$ for case (i), or $d^4n/(dzdm_1dqde)$ for cases (ii) and (iii). 
In {\it Model-delayed}, all subsets contribute to the GWB, and we therefore write $h_c^2=h_{c,\rm{star}}^2+h_{c,\rm{Tr}}^2+h_{c,\rm{Tr-ej}}^2$, where $h_{c,\rm{star}}^2$ is computed via equation (\ref{hcdE}) 
and  $h_{c,\rm{Tr}}^2$, $h_{c,\rm{Tr-ej}}^2$ are obtained via equation (\ref{eq:hcanalytic}). 
In {\it  Model-stalled}, only triple interactions can drive MBHB coalescences; the GWB is therefore computed as  $h_c^2=h_{c,\rm{Tr}}^2+h_{c,\rm{Tr-ej}}^2$, where both terms are obtained via equation (\ref{eq:hcanalytic}).

%%%%%%%%%%%%%%%%%%%%%%%%%%%%%%%%%%%%%%%%%%%%%%%%%%%%%%%%%%%%%%%%%%%%%%%%%%%%%%%%%%%%%
%%%%%%%%%%%%%%%%%%%%%%%%%%%%%%%%%%%%%%%%%%%%%%%%%%%%%%%%%%%%%%%%%%%%%%%%%%%%%%%%%%%%%
\section{Results}
\label{sec:results}

%%%%%%%%%%%%%%%%%%%%%%%%%%%%%%%%%%%%%%%%%%%%%%%%%%%%%%%%%%%%%%%%%%%%%%%%%%%%%%%%%%%%%
\subsection{MBH merger rates}

%%%%%%%%%%%%%%%%%%%%%%%%%%%%%%%%%%%%%%%%%
\begin{figure*}
	\includegraphics[scale=0.43,clip=true]{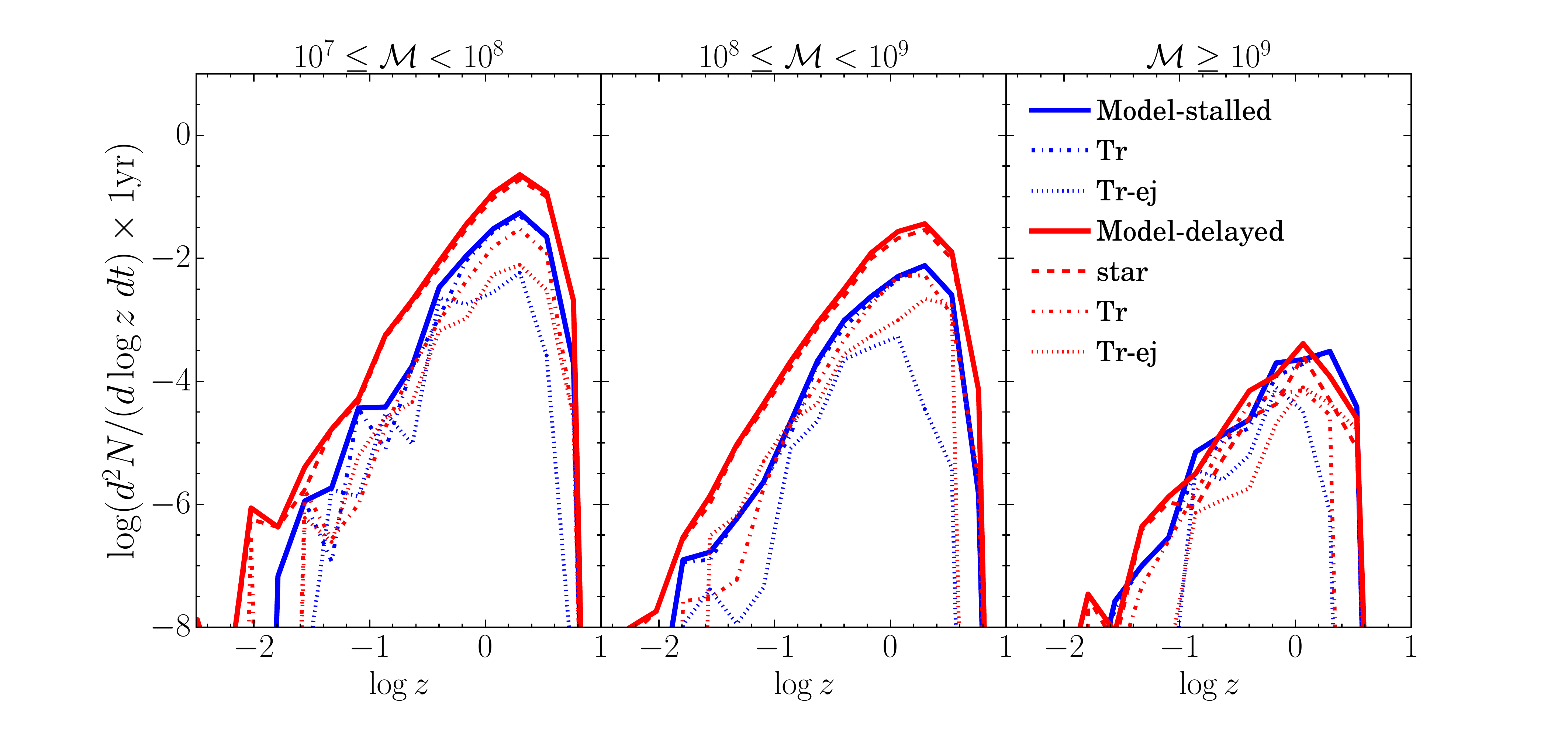}
	\caption{Redshift distribution of merging MBHBs in different chirp mass bins, as labelled at the top of the right panel. Line styles are described in the figure.}
	\label{fig:zdistmcut}
\end{figure*}
%%%%%%%%%%%%%%%%%%%%%%%%%%%%%%%%%%%%%%%%%

We first focus on the MBHB merger rates predicted by {\it Model-delayed} and {\it Model-stalled}. Being interested primarily in the PTA signal, we concentrate on 
systems with ${\cal M}>10^{7}\msun$. Results are reported for both {\it HS} and {\it LS} models in table \ref{tab2} and plotted for the {\it LS} model (our default choice) in figure \ref{fig:zdistmcut}. 
The table lists the total merger rates for selected chirp-mass ranges, and highlights the relative contributions of different MBHB sub-populations, providing a number of useful information:

%%%%%%%%%%%%%%%%%%%%%%%%%
\begin{table*}
	\centering
	\begin{tabular}{c|ccc|cc|ccc|cc}
		\hline
		\multicolumn{11}{c}{LS}\\
		\hline
		\multirow{3}{*}{$\mathcal{M}$ [$\rm M_{\odot}$]}&\multicolumn{5}{c|}{{\it Model-delayed}}&\multicolumn{5}{c}{{\it Model-stalled}}\\
		&\multirow{2}{*}{Rate [yr$^{-1}$]}&\multicolumn{2}{c|}{binaries}&\multicolumn{2}{c|}{triplets}&\multirow{2}{*}{Rate [yr$^{-1}$]}&\multicolumn{2}{c|}{binaries}&\multicolumn{2}{c}{triplets}\\  
		&& star &  gas & Tr (quad) & Tr-ej && star &  gas & Tr (quad) & Tr-ej \\
		\hline
		$10^{7}-10^{8}$ & 0.118 & 80.0\% & 5.1\% & 12.4\%(1.3\%) & 3.6\% & 0.028 & -- & -- & 89.3\%(77.3\%) & 10.7\% \\
		$10^{8}-10^{9}$ & 0.022 & 75.9\% & 2.9\% & 15.0\%(3.4\%) & 6.2\% & 4.4$\times 10^{-3}$ & -- & -- & 93.6\%(81.8\%) & 6.3\% \\
		$10^{9}$ & 1.8$\times 10^{-4}$ & 50.7\% & 0.04\% & 28.0\%(11.2\%) & 21.3\% & 1.9$\times 10^{-4}$ & -- & -- & 84.6\%(77.4\%) & 15.4\% \\
		\hline
		\multicolumn{11}{c}{HS}\\
		\hline
		\multirow{3}{*}{$\mathcal{M}$ [$\rm M_{\odot}$]}&\multicolumn{5}{c|}{{\it Model-delayed}}&\multicolumn{5}{c}{{\it Model-stalled}}\\
		&\multirow{2}{*}{Rate [yr$^{-1}$]}&\multicolumn{2}{c|}{binaries}&\multicolumn{2}{c|}{triplets}&\multirow{2}{*}{Rate [yr$^{-1}$]}&\multicolumn{2}{c|}{binaries}&\multicolumn{2}{c}{triplets}\\  
		&& star &  gas & Tr (quad) & Tr-ej && star &  gas & Tr (quad) & Tr-ej \\
		\hline
		$10^{7}-10^{8}$ & 0.079 & 79.2\% & 4.2\% & 13.0\%(2.2\%) & 3.6\% & 0.044 & -- & -- & 89.3\%(64.7\%) & 10.7\% \\
		$10^{8}-10^{9}$ & 0.020 & 76.7\% & 4.3\% & 15.1\%(3.5\%) & 3.9\% & 5.4$\times 10^{-3}$ & -- & -- & 86.6\%(65.8\%) & 13.4\% \\
		$10^{9}$ & 2.4$\times 10^{-4}$ & 72.2\% & 2.1\% & 21.3\%(4.8\%) & 4.5\% & 1.9$\times 10^{-4}$ & -- & -- & 79.9\%(65.6\%) & 20.1\% \\
		\hline
	\end{tabular}
	\caption{Merger rate and population composition of the MBHBs with chirp mass in the three most massive mass bins. The number in parenthesis refer to the fraction of prompt mergers originated by a quadruple system.}
	\label{tab2}
\end{table*}
%%%%%%%%%%%%%%%%%%%%%%%%%%

%%%%%%%%%%%%%%%%%%%%
\begin{enumerate}
\item As anticipated, there is little difference between the {\it HS} and {\it LS} models; rates are very similar for all mass ranges and so are the fractions of mergers due to individual channels.
\item The only exception is the fraction of mergers due to quadruple interactions, which is larger in the {\it LS} model. This is due to this model's large occupation fraction of MBHs with $M<10^{5}\msun$, 
which is also responsible for the presence of very low mass ratio binaries (DB17), which are absent in the {\it HS} model. 
We stress, however, that these low mass ratio systems have little effect on the level of the GWB.
\item In {\it Model-delayed}, more than $75\%$ of all merging systems with ${\cal M}<10^9\msun$ are `regular' binaries, the vast majority of which reside in gas poor environments. 
This validates our assumption that regular binaries evolve via stellar hardening only.
\item The importance of triple interactions is a monotonically increasing function of mass. For  ${\cal M}>10^9\msun$, about half of the mergers are due to this channel. 
This is because very massive galaxies experience several mergers in their lifetime, hence typical MBHB merger timescales are longer than the time occurring between subsequent galaxy mergers. 
{\it Model-delayed} is therefore similar to {\it Model-stalled} at such high masses, but we checked that systems with ${\cal M}>10^9\msun$ (which are quite rare, see the `Rate' column in the table) 
contribute less than $10\%$ to the overall GWB signal.
\item For  ${\cal M}<10^9\msun$, merger rates of {\it Model-delayed} are about four times higher than those of {\it Model-stalled}. Triple interactions have therefore limited efficiency (about $30\%$) in solving the final-parsec problem, since many of them simply end up with the ejection of one of the MBHs (about $70\%$) without the left-over binary merging. 
Note that, conversely, rates are comparable for ${\cal M}>10^9\msun$, for the reasons discussed in the previous point.
\item In general,  $\approx 80-90\%$ of triplet-induced mergers are prompt (Tr), whereas only about $10-20\%$ are due to GW hardening following the ejection of the lighter of the three MBHs (Tr-ej). 
This is true for both {\it Model-delayed} and {\it Model-stalled}.
\end{enumerate}
%%%%%%%%%%%%%%%%%%%%%%
%
The redshift distribution of merging systems is shown in figure \ref{fig:zdistmcut} for the LS model. Counter intuitively, mergers do not appear to be shifted, on average, to lower redshifts in {\it Model-stalled}. 
This is likely because typical MBHB merger timescales in {\it Model-delayed} are of the order of Gyrs, especially when systems are stellar driven (which are the vast majority at  ${\cal M}>10^7\msun$), similar to the timescale of subsequent mergers that trigger triple interaction.  
Within the triple-induced mergers, however, the Tr-ej sub-group (dotted curves in figure) tends to coalesce at lower redshifts than the Tr one (dashed curves in the figure). 
This is because the former is comprised of left-over systems that merge because of GW emission only, and 
their coalescence timescale is skewed towards values of several Gyrs (cf Paper II, figure 7), thus shifting the peak of the merger rate to lower $z$. The Tr binaries, conversely, 
typically coalesce in few hundred Myrs. 
The overall shape and normalisation of the rates are in line with estimates from other authors \citep[e.g.,][]{Blecha2016}, and the implied total merger rate of MBHBs with ${\cal M}>10^7\msun$ 
is about 0.14 yr$^{-1}$ in {\it Model-delayed} and  0.032 yr$^{-1}$ in {\it Model-stalled}.

%%%%%%%%%%%%%%%%%%%%%%%%%%%%%%%%%%%%%%%%%%%%%%%%%%%%%%%%%%%%%%%%%%%%%%%%%%%%%%%%%%%%%
%%%%%%%%%%%%%%%%%%%%%%%%%%%%%%%%%%%%%%%%%%%%%%%%%%%%%%%%%%%%%%%%%%%%%%%%%%%%%%%%%%%%%
\subsection{Stochastic GW background}

%%%%%%%%%%%%%%%%%%%%%%%%%%%%%%%%%%%%%%%%%
\begin{figure*}
 \begin{minipage}[b]{0.47\textwidth}
   \centering
   \includegraphics[scale=0.43]{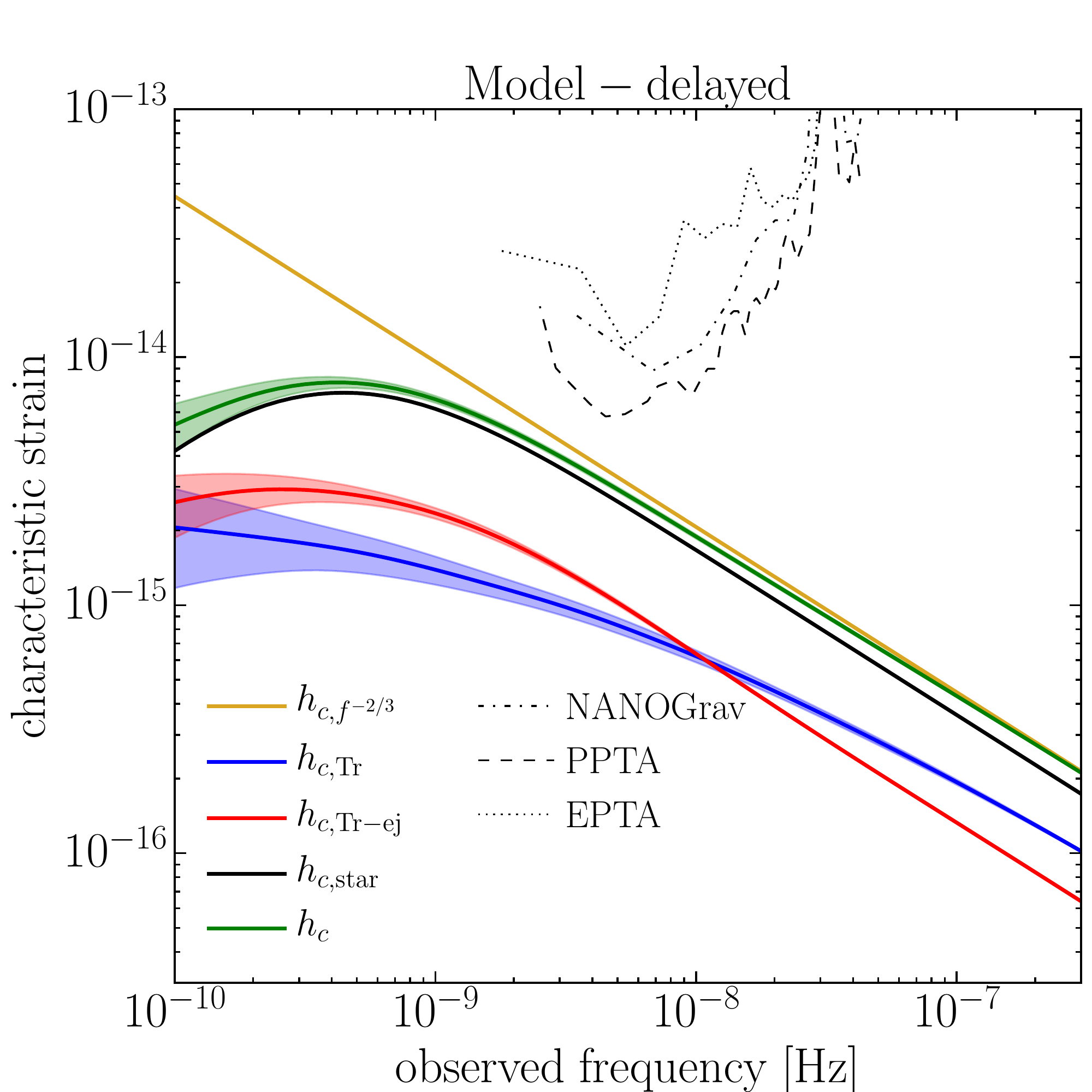}
 \end{minipage}
 \ \hspace{1mm} \
 \begin{minipage}[b]{0.47\textwidth}
  \centering
   \includegraphics[scale=0.43]{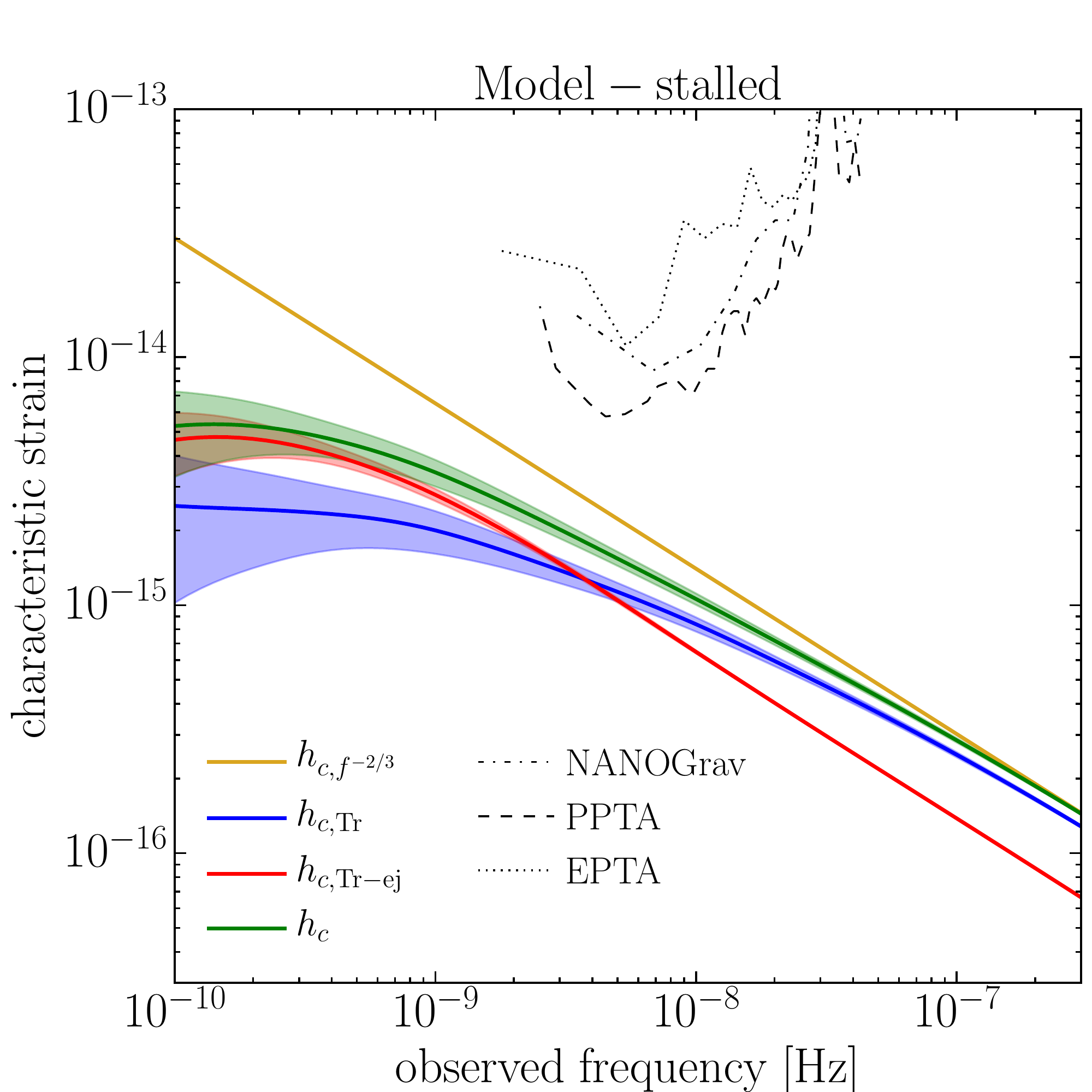}
 \end{minipage}
 \caption{Stochastic GWB from MBHBs, divided in each of the contributing components, as indicated in each panel. Solid lines show the mean values of
 the characteristic strain $h_c$ of each component, whereas the shaded area marks the standard deviation inferred from 100 Monte Carlo simulations of the MBHB population. Current sensitivities of 
 major PTA experiments are also shown.}
\label{fig:hc2panel}
\end{figure*}
%%%%%%%%%%%%%%%%%%%%%%%%%%%%%%%%%%%%%%%%%

%%%%%%%%%%%%%%%%%%%%%%%%%%%%%%%%%%%%%%%%%
\begin{figure*}
 \begin{minipage}{0.47\textwidth}
	\includegraphics[scale=0.43,clip=true]{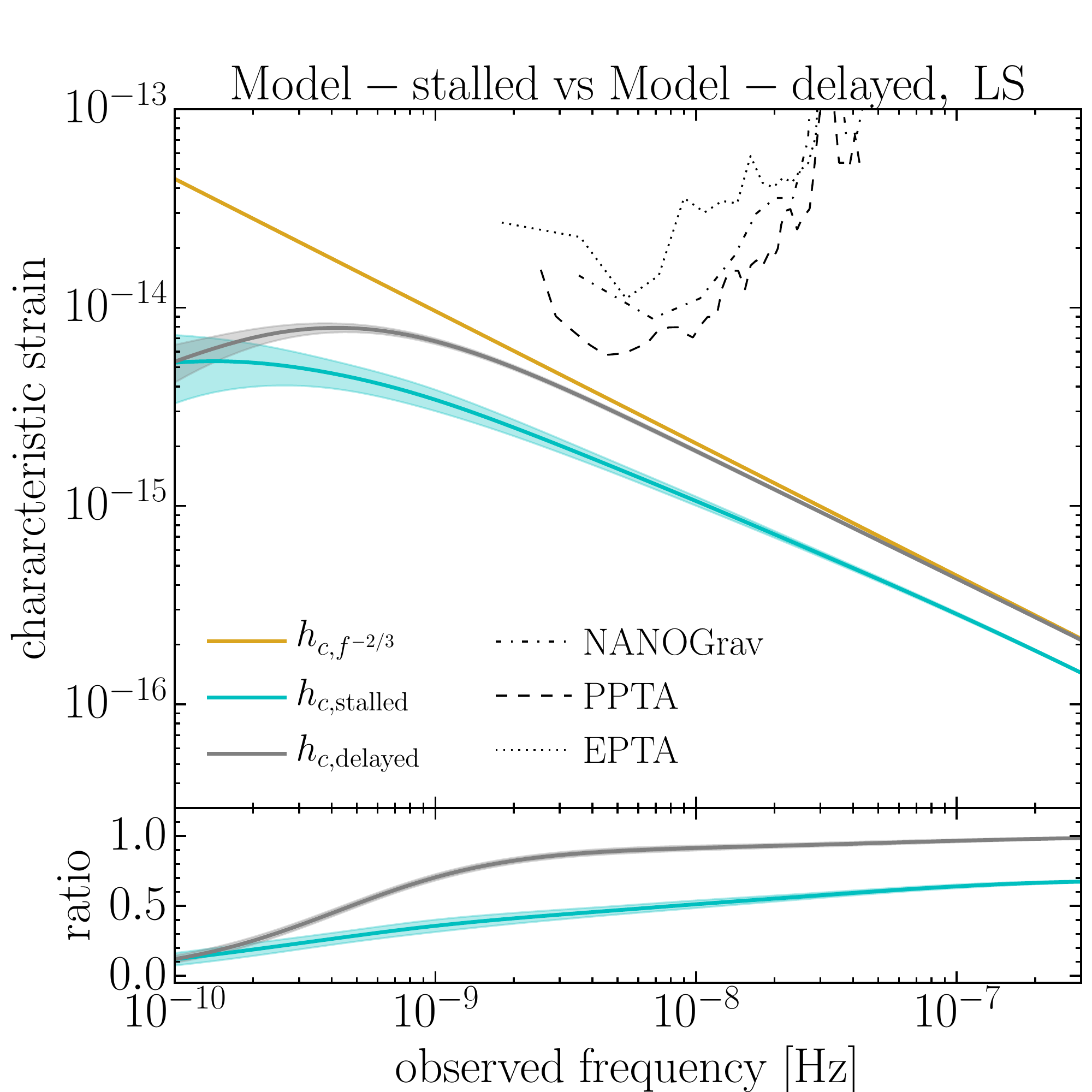}
 \end{minipage}
 \ \hspace{1mm} \
 \begin{minipage}{0.47\textwidth}
	\includegraphics[scale=0.43,clip=true]{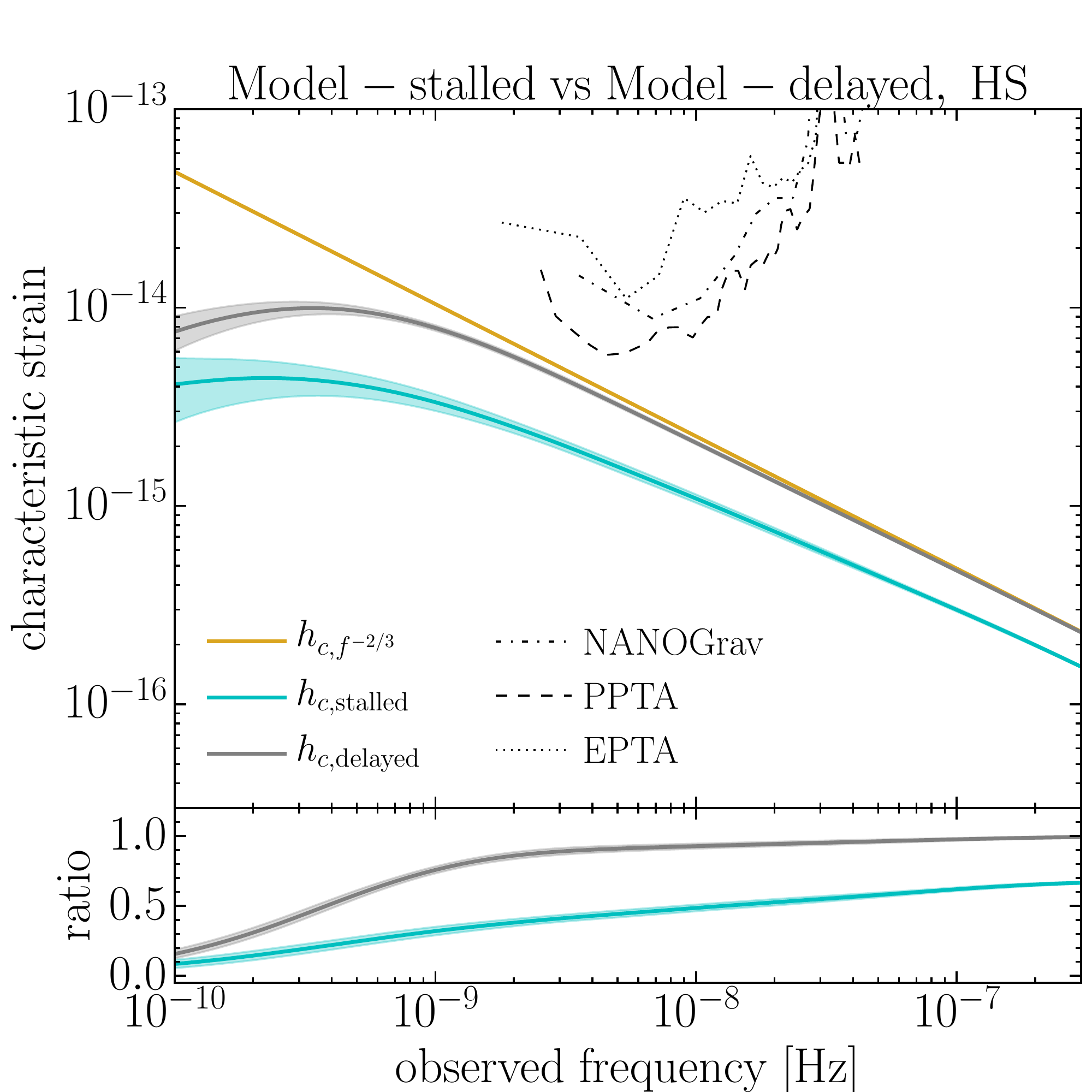}
 \end{minipage}

 \caption{Comparison of the stochastic  GWB generated by  {\it Model-delayed} (grey) and {\it Model-stalled} (light-blue). Lines and shaded areas have the same meaning as in figure \ref{fig:hc2panel}. 
 The bottom inset shows the ratio between either models and a reference $f^{-2/3}$ power law generated by a GW-driven population of circular binaries, shown as an orange line in the main plots. Here, we present 
 results for both LS (left) and HS (right) seed models.}
\label{fig:hccomp}
\end{figure*}
%%%%%%%%%%%%%%%%%%%%%%%%%%%%%%%%%%%%%%%%%

Figure \ref{fig:hc2panel} shows the stochastic GWB produced by the two models. The figure is obtained by combining 100 Monte Carlo sampling of the $d^4n/(dzdm_1dqde)$ distribution.
To assess the overall effect of MBHB stalling in the normalisation of the expected GWB, we first ignore effects due to stellar hardening and eccentricity and compute the GWB as due to circular GW-driven binaries. In this case, standard calculations  
gives $h_c=A[f/(1 \ {\rm yr}^{-1})]^{-2/3}$, shown as a golden line in figure~\ref{fig:hc2panel}. In practice, we evaluate the integrals in eq.~\ref{hcdE} under the assumption that $df_r/dt_r$ is solely determined by the (circular) emission of GWs (i.e., we set $\mathcal{A} = 0$ in eq.~\ref{eq:f_evol}). We find $A=1\times10^{-15}$ and $A=0.7\times 10^{-15}$ for {\it Model-delayed} and {\it Model-stalled} respectively.

In {\it Model-delayed}, the vast majority of the GWB ($h_{c,{\rm star}}\approx 0.9 h_c$) is produced by regular binaries evolving via stellar hardening, with triplet-induced mergers (either 
prompt or following ejection) playing a sub-dominant role. As long as MBHBs are not highly eccentric, the spectral turnover is at $f<1$ nHz (black and green lines in the left panel of figure \ref{fig:hc2panel}), 
and the signal only mildly departs from the $f^{-2/3}$ power law at frequencies relevant to PTA detection. 
This is certainly true in our model, which assumes circular binaries, but the shape of the spectrum is hardly affected by eccentricity up to $e\approx 0.5$ \citep{Sesana2015b,Taylor2017,Chen2017}. 
We note, however, that the evolution of the eccentricity of stellar driven binaries strongly depends on its initial value at binary formation, which is a poorly understood parameter, with N-body simulations of 
merging galactic bulges resulting in a wide range of MBHB eccentricities \citep[see][for a review]{Dotti2012}. Conversely, the signal in {\it Model-stalled} already departs from the $f^{-2/3}$ power law 
at $f\approx 3\times 10^{-8}$ Hz, and at $1$ nHz it is already a factor $\sim 2$ below its nominal $f^{-2/3}$ value. This is due to the non-negligible eccentricity of MBHBs merging via triple interactions. 
Unlike in the stellar hardening scenario, the presence of binaries with high eccentricities is not just a possibility in this case, but an {\it inevitable} outcome of the three-body MBH dynamics (cf Paper I and Paper II). 
Note the relative contribution of promptly induced coalescences ($h_{c,\rm{Tr}}$) and GW-driven coalescences following ejection of one of the triplet's members ($h_{c,\rm{Tr-ej}}$). 
The normalisation of the latter contribution is a factor $\sim 2$ lower, being Tr-ej systems about $20\%$ of the overall triplet-induced coalescences. However, the contributions of the two sub-populations 
have different spectral shapes, crossing at $f\approx 3\times 10^{-9}$Hz, below which Tr-ej becomes dominant. This is due to the different eccentricity distribution of the two sub-populations, 
as we will see in the next subsection. 

Figure \ref{fig:hccomp} visualises the difference between the simple $f^{-2/3}$ power law, {\it Model-delayed} and {\it Model-stalled}. Results are shown for both {\it LS} and {\it HS} models, 
to stress their similarity. The plot clearly shows that {\it Model-delayed} closely follows the simple power law model at least down to $2\times 10^{-9}$Hz, with a low-frequency drop due to stellar driven evolution. 
The ratio between  {\it Model-stalled} and the single power-law model, as already mentioned, is about 0.7 at high frequencies, monotonically decreasing to about 0.1 at 0.1 nHz. 
Compared to {\it Model-delayed}, {\it Model-stalled} produces a GWB that is a fraction 2-to-3 smaller in the frequency range 1-10 nHz, most relevant to PTA experiments. The result holds for both {\it LS} and {\it HS} models,
with minimal differences.

%%%%%%%%%%%%%%%%%%%%%%%%%%%%%%%%%%%%%%%%%%%%%%%%%%%%%%%%%%%%%%%%%%%%%%%%%%%%%%%%%%%%%
\subsubsection{Eccentricity distribution}

%%%%%%%%%%%%%%%%%%%%%%%%%%%%%%%%%%%%%%%%%
\begin{figure*}
 \begin{minipage}[b]{0.47\textwidth}
   \centering
   \includegraphics[scale=0.3]{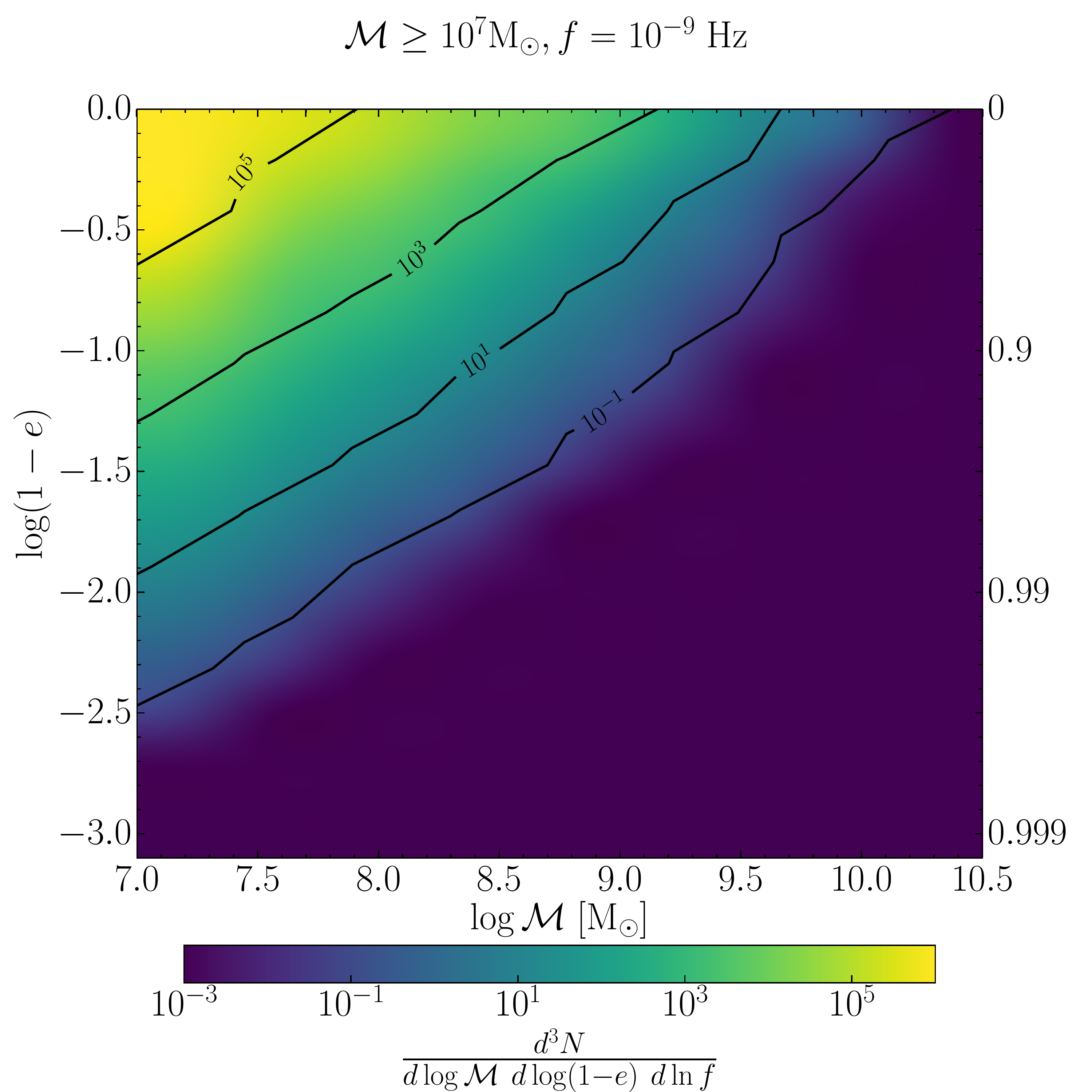}
 \end{minipage}
 \ \hspace{3mm} \
 \begin{minipage}[b]{0.47\textwidth}
  \centering
   \includegraphics[scale=0.3]{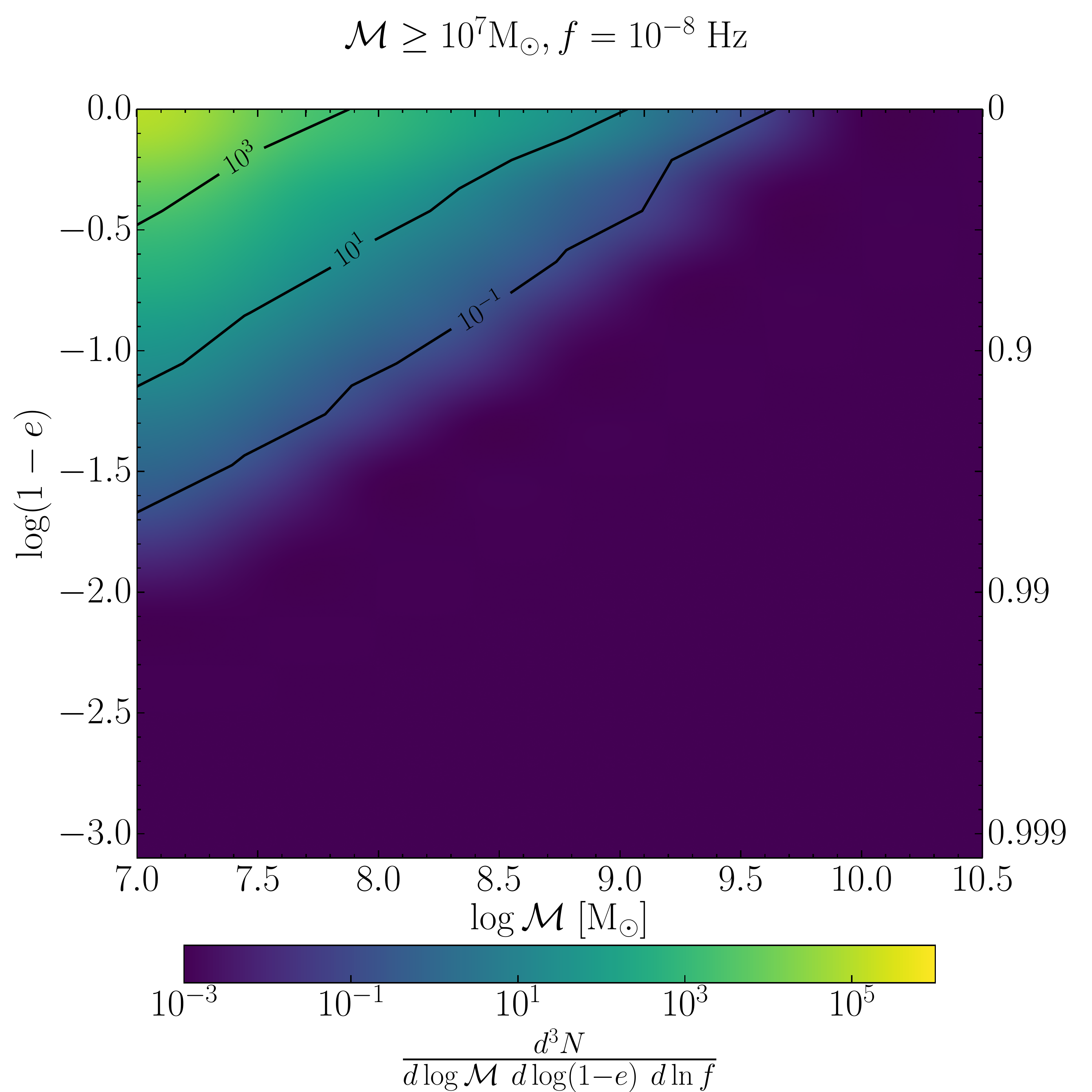}
 \end{minipage}
 
 \begin{minipage}[b]{0.47\textwidth}
   \centering
   \includegraphics[scale=0.17]{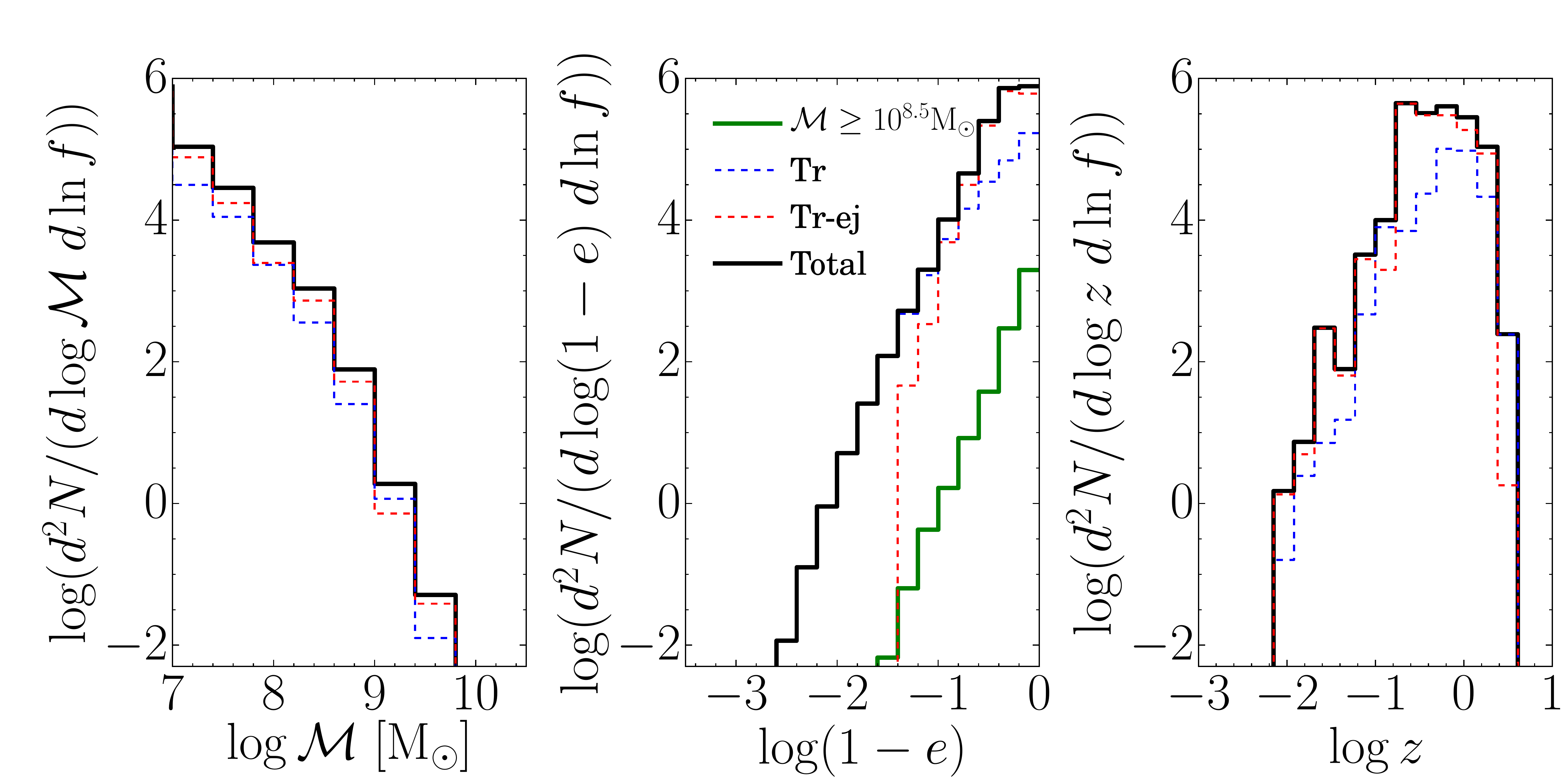}
 \end{minipage}
 \ \hspace{3mm} \
 \begin{minipage}[b]{0.47\textwidth}
  \centering
   \includegraphics[scale=0.17]{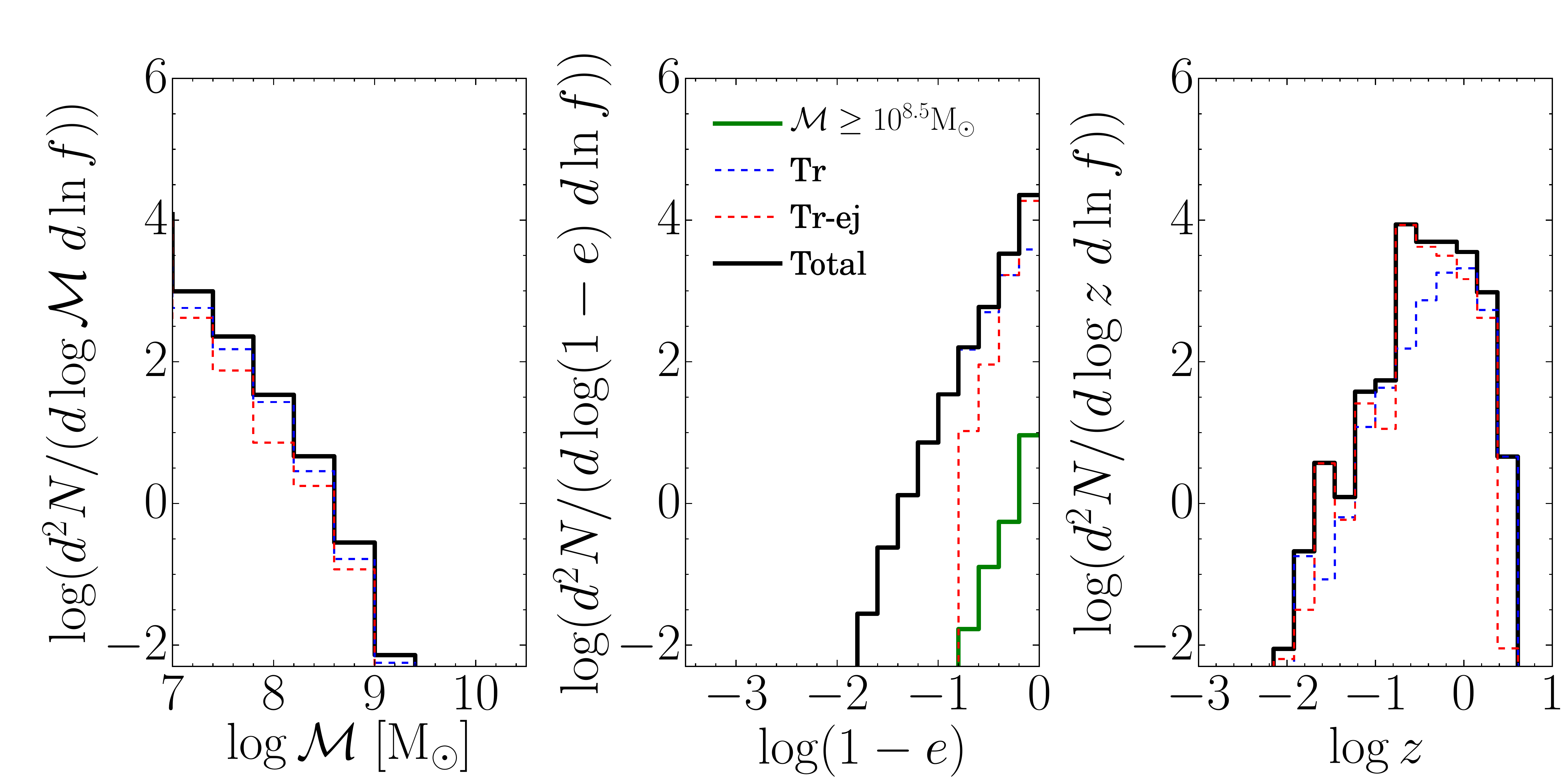}
 \end{minipage}
 \caption{Properties of individual MBHBs contributing to the GW signal within an observed orbital frequency $\Delta{f}=f$ around $f=1$ nHz (left) and $f=10$ nHz (right). 
The upper panels show the differential distribution of sources in the chirp mass -- circularity plane (with the eccentricity $e$ shown on the right of each figure). 
The lower panels show the marginalised distributions of the number of sources as a function of chirp mass (left), circularity (centre) and redshift (right). The 
legend of the histogram linestyles is shown in the central panels.}
\label{fig:dN_dMdzdc_f}
\end{figure*}
%%%%%%%%%%%%%%%%%%%%%%%%%%%%%%%%%%%%%%%%%

As mentioned above, the occurrence of high eccentricities is a critical and inevitable feature of triple-induced inspirals. 
This might be relevant for the detection of individually resolvable binaries, for which eccentric templates might be necessary \citep{Taylor2016}. 
To investigate the actual distribution of individual MBHB eccentricities, we need to convert the merger number densities in equation (\ref{eq:hcanalytic}) 
into the instantaneous number of systems in the sky at a given frequency, i.e.
\begin{equation}
\frac{d^5N}{dzdm_1dqded{\rm ln}f}=\frac{d^4n}{dzdm_1dqde'}\frac{dV_c}{dz}\frac{dz}{dt}\frac{dt}{d{\rm ln}f},
\end{equation}
where $e'$ is the eccentricity computed at $100R_G$, $V_c$ is the comoving volume, $dV_c/dz$, $dz/dt$ are known once a cosmology is assumed \citep{Hogg1999}, and
\begin{align}
  \frac{dt}{d{\rm ln}f} & = \frac{5}{96} (2\pi)^{-8/3} \mathcal{M}^{-5/3} f^{-8/3} F(e)^{-1} 
  \\
F(e) & = \frac{1+(73/24)e^2+(37/96)e^4}{(1-e^2)^{7/2}}.
\label{eq:dfdt}
\end{align}
The eccentricity $e$ at the desired frequency $f$ is obtained by evolving $e'$ backwards from $100R_G$ to $f$ by 
using the standard evolution of eccentric binaries in the quadrupole approximation \citep{Peters1963,Chen2017}. 
Note that $F(e)$ is a strong increasing function of $e$, thus $dt/d{\rm ln}f$ is much shorter for very eccentric binaries, down-weighing their relative number at a given observed frequency. 

The distribution of the number of emitting binaries, integrated in redshift, in the 
circularity\footnote{The circularity is defined as 1-$e$. Its logarithm is often used for plotting purposes, to highlight tails of high eccentricities.}-mass plane is shown in figure \ref{fig:dN_dMdzdc_f} 
at two observed {\it orbital} reference frequencies $f=1$nHz and $f=10$nHz. The number of sources is obviously dominated by low mass MBHBs, with a long tail of few sources extending up to ${\cal M}=10^{10}\msun$. 
The overall distribution is dominated by light, rather circular binaries, but possible eccentricities extend up to $e>0.99$.

Marginalised source distributions are shown in the lower panel as a function of chirp mass, circularity and redshift. As expected, the number of sources is dominated by low-mass systems (which, however, do not 
contribute much to the total GWB) and the redshift distribution peaks around $z \sim 1$, consistent with, e.g, \citet{Sesana2009} and \citet{Blecha2016}. As already noticed, the Tr and Tr-ej sub-populations 
behave quite differently, the latter peaking at lower redshifts (because of the long coalescence timescales). Moreover, the circularity distributions are also distinct: 
the Tr population extends to $1-e\approx 0.003(0.03)$ at $f=1(10)$ nHz, whereas the the Tr-ej populations hardly goes below $1-e\approx 0.1$, preferentially selecting rather circular binaries. 
This can be understood by looking at figure 10 of Paper II. Prompt coalescences (i.e. the Tr population) are generally caused by the formation of either a temporarily highly eccentric binary (mostly as a result of 
secular Kozai-Lidov oscillations), or a compact system of moderate eccentricity (in the case of chaotic energy and angular momentum exchanges). 
The resulting eccentricity distribution is therefore extremely broad, spanning more than six orders of magnitude at the innermost circular stable orbit. 
Conversely, coalescences driven by GW emission after an ejection (i.e. the Tr-ej population) preferentially select the systems that did not reach sufficiently high eccentricities 
 to promptly coalesce, but which are still sufficiently eccentric that their coalescence time is shorter than the Hubble time. 
 The result is a much narrower (and mass dependent) allowed eccentricity range, which does not reach the high values of the Tr population. 
 The green curves in the two central panels select binaries with ${\cal M}>10^{8.5}\msun$, which are the loudest GW sources and which are thus more likely to be individually resolved. 
 Although the distribution favours circular binaries, ${\cal O}(10)$ systems have eccentricities higher than 0.7 at $f\approx 1$ nHz. Eccentric resolvable sources are therefore a rather common occurrence 
 if merging MBHBs are mostly driven by triple interactions.

%%%%%%%%%%%%%%%%%%%%%%%%%%%%%%%%%%%%%%%%%%%%%%%%%%%%%%%%%%%%%%%%%%%%%%%%%%%%%%%%%%%%%
\subsection{A realistic lower bound to the GWB: implications for PTA detectability}

Our results imply that triple interactions efficiently counteract the effect of stalling, driving MBHBs to coalescence and resulting in a GWB being reduced by a factor of 2-to-3 only in the PTA band. We can now use this fact to derive a robust lower limit to the amplitude distribution of the expected GWB, 
based on our best astrophysical knowledge of MBH assembly and dynamics following galaxy mergers. 

Our scope is not to make realistic predictions of time to detection \citep[see, e.g.][]{Siemens2013,Taylor2016b}, but to assess the impact of the most pessimistic scenario 
on GWB detectability by PTAs. As such, we consider a simple model based on the following assumptions:

%%%%%%%%%%%%%%%%%%%%%%%%%%%%%%%%%%%%%%%%%
\begin{figure}
	\includegraphics[scale=0.42,clip=true]{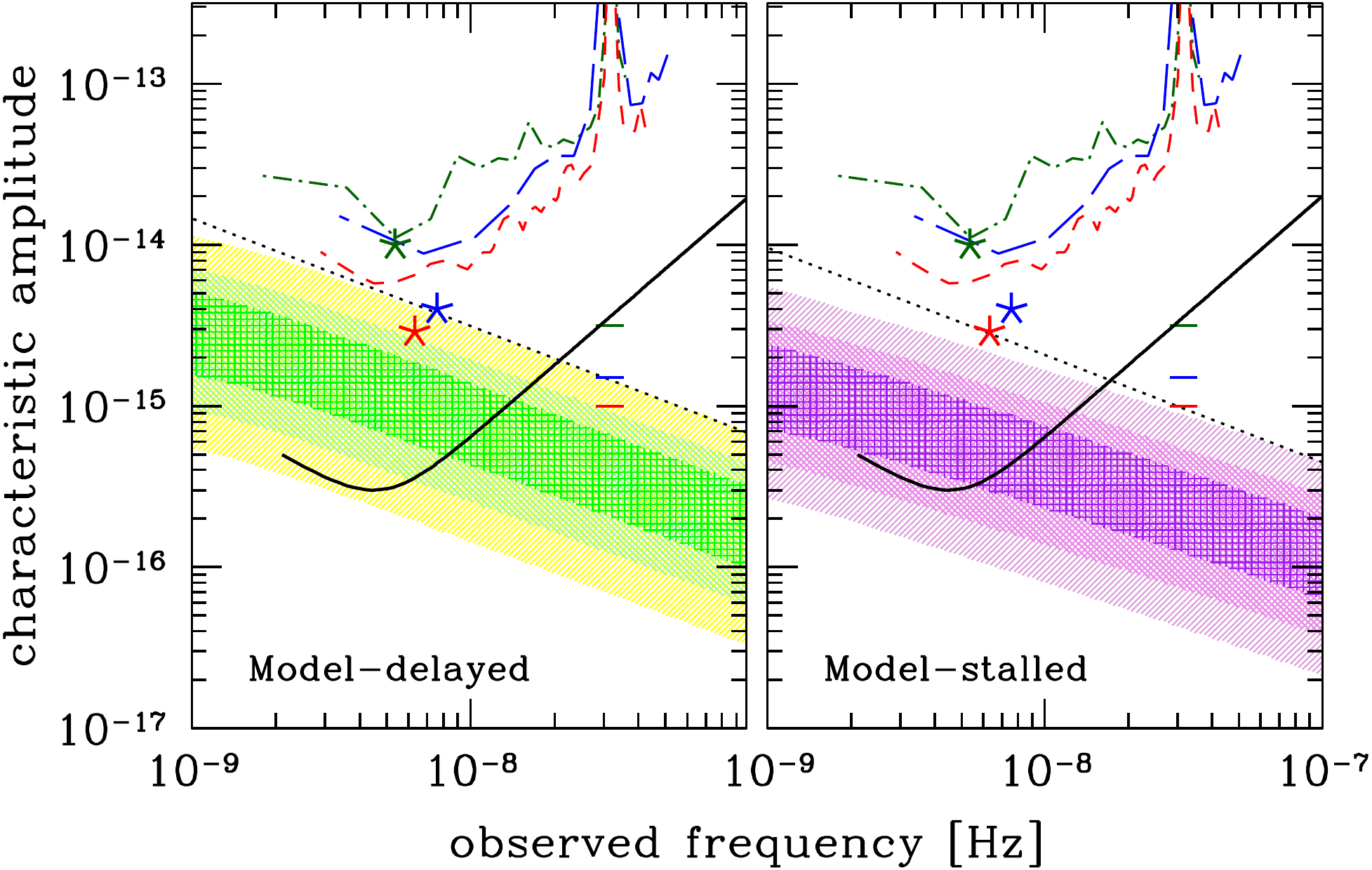}
	\caption{$h_c$ vs $f$ for {\it Model-delayed} (left panel) and {\it Model-stalled} (right panel) assuming the conservative MBH population model from \protect\cite{Shankar2016}. 
		In each panel, the shaded areas represent the $68\%$, $95\%$ and $99.7\%$ confidence intervals of the signal amplitude. The jagged curves are current PTA sensitivities: EPTA (dot-dashed green), NANOGrav (long-dashed blue), 
		and PPTA (short-dashed red). For each sensitivity curve, stars represent the integrated upper limits to an $f^{-2/3}$ background, and the horizontal ticks are their extrapolation at $f=1$ yr$^{-1}$. 
		The solid black line represent the typical sensitivity level of a conservative SKA-type array formed by 50 pulsars monitored at 100ns precision for 15 years. A dotted black line with slope $f^{-2/3}$ 
		is also added to guide the eye.}
	\label{fig:hc_shank}
\end{figure}
%%%%%%%%%%%%%%%%%%%%%%%%%%%%%%%%%%%%%%%%%

%%%%%%%%%%%%%%%%%%%%%
\begin{enumerate}
\item We take a GWB amplitude distribution predicted by a MBHB population model based on the MBH-galaxy scaling relations proposed by \citet{Shankar2016}, which are based on a putative observational selection bias on the resolvability of the MBH influence sphere.
This choice is solely based on the fact that those relations provide a very conservative estimate of the stochastic GWB. In fact, \citet{Sesana2016} showed that 
they result in a $95\%$ confidence GWB amplitude distribution in the range $1.4\times10^{-16} < A < 1.1\times10^{-15}$, with a median value of $A=4\times10^{-16}$, well below current PTA limits. Also, consistently with this choice, the semianalytic galaxy-formation model utilised in this paper reproduces the scaling relations of \citet{Shankar2016} \citep[cf.][]{Barausse2017}.
\item We draw $A$ from this distribution and apply a correction factor $C(f)_{\rm delayed}=h_{c,{\rm delayed}}(f)/h_{c,f^{2/3}}(f)$ and $C(f)_{\rm stalled}=h_{c,{\rm stalled}}(f)/h_{c,f^{2/3}}(f)$, 
shown in the lower panels of figure \ref{fig:hccomp}. In both cases, the GWB then takes the form $h_{c,X}=C(f)_X\times A(f/1 \ {\rm yr}^{-1})^{-2/3}$.
\item We consider an idealised SKA-type PTA. Following \citet{Janssen2015}, we make the conservative assumption that SKA will be able to monitor up to 50 MSPs with rms precision of 100ns. 
We then explore the detection probability (DP) as a function of observing time $T$ and number of pulsars $N_p$ in the array.  
\end{enumerate}
%%%%%%%%%%%%%%%%%%%%%%%%
%
In particular, assumptions (i) and (ii) provide a realistic projection of how low the GWB can get, by combining the existence of moderately light MBHBs to stalling.

%%%%%%%%%%%%%%%%%%%%%%%%%%%%%%%%%%%%%%%%%
\begin{figure*}
	\begin{tabular}{cc}
		\includegraphics[scale=0.35,clip=true]{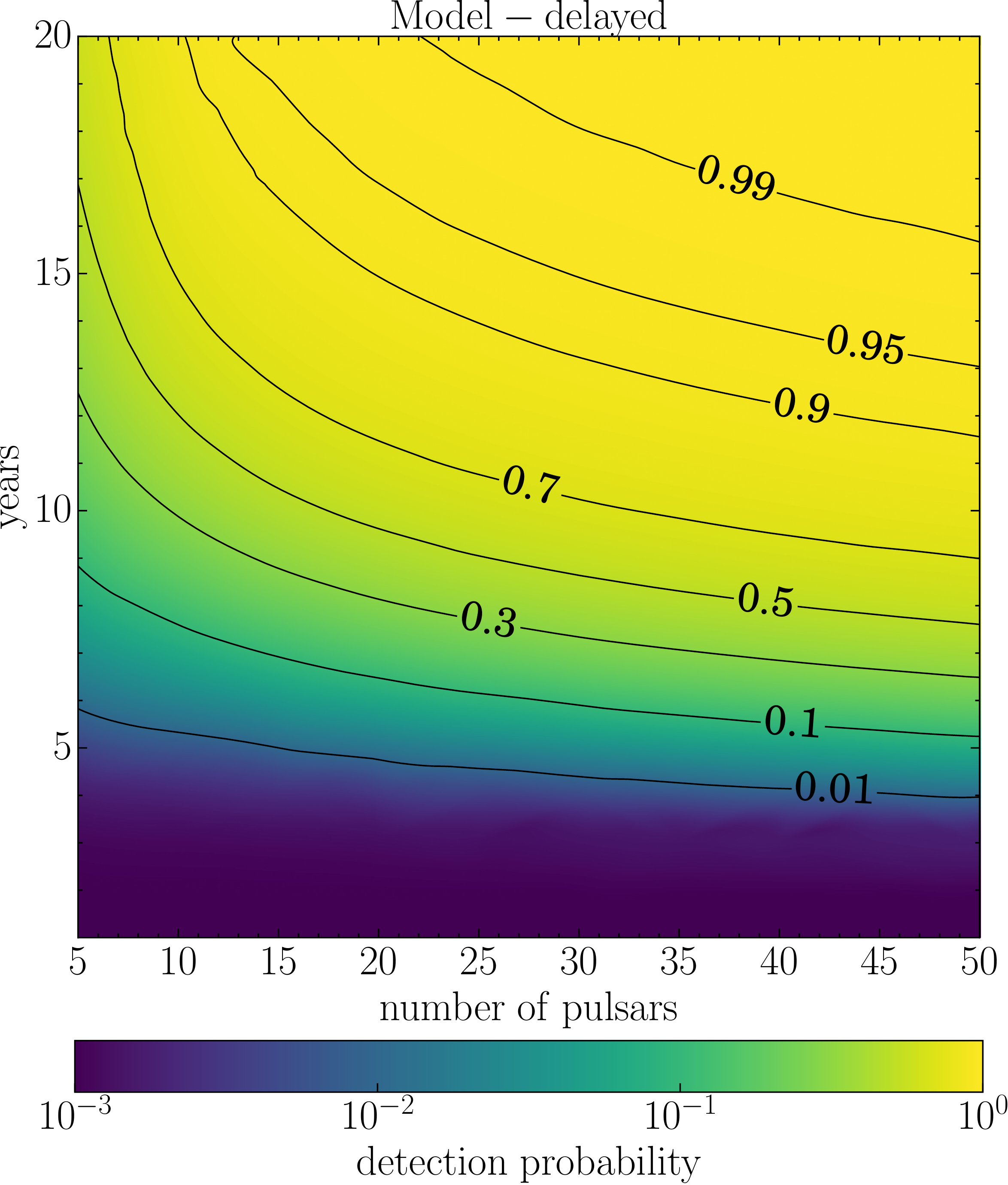}&
		\includegraphics[scale=0.35,clip=true]{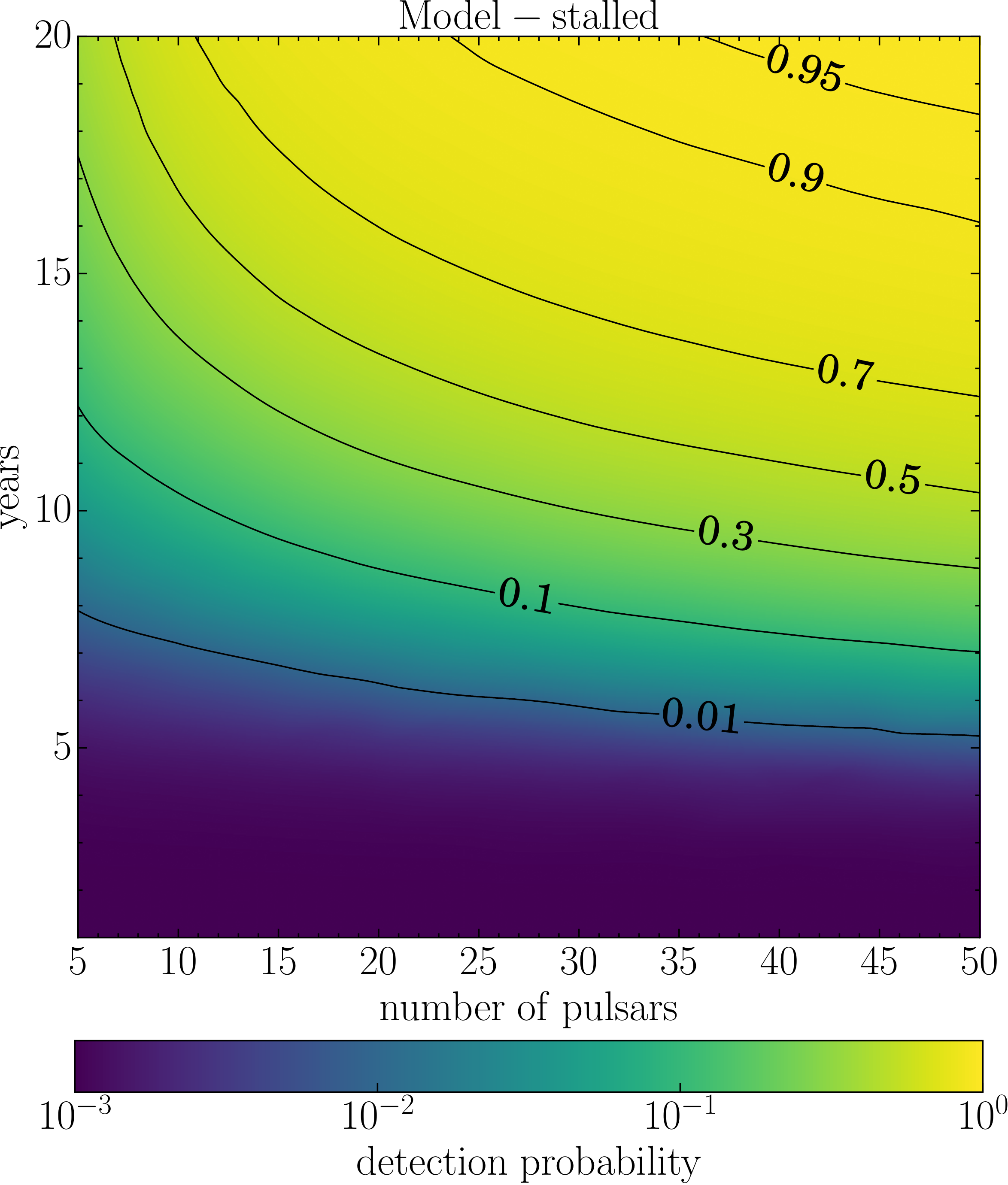}\\
	\end{tabular}
	\caption{Detection probability contour plot in the number of pulsar vs observation time plane. All pulsars are assumed to have an equal rms residual of 100ns.}
	\label{fig:dp_shank}
\end{figure*}
%%%%%%%%%%%%%%%%%%%%%%%%%%%%%%%%%%%%%%%%%

PTA detectability under assumption (iii) is computed using the framework developed by \citet{Rosado2015}. If the noise is described by a Gaussian distribution with zero mean and variance $\sigma_0$, 
the DP of a stochastic signal is also described by a Gaussian distribution with mean $\mu_1$ and dispersion $\sigma_1$ given by

%%%%%%%%%%%%%%%%%%%%%%%
\begin{equation}
\label{eq:DP}
\gamma_\text{B}=\frac{1}{2}\text{erfc}\left( \frac{\sqrt{2}\sigma_0\text{erfc}^{-1}(2\alpha_0)-\mu_1}{\sqrt{2}\sigma_1}\right),
\end{equation}
%%%%%%%%%%%%%%%%%%%%%%
%
where erfc is the complementary error function and $\alpha_0$ is the chosen false alarm probability threshold, which we fix at 0.001. Under our simplifying assumptions of an array of equal pulsars, 
randomly distributed in the sky and monitored for the same timespan $T$, $\mu_1$  $\sigma_0$ and $\sigma_1$ take the form

%%%%%%%%%%%%%%%%%%%%
\begin{align}
\label{eq:mu1B}
\mu_1&=N_p(N_p-1)T\int df \frac{\Gamma^2 S_h^2}{(P+S_h)^2+\Gamma^2S_{h}^2}, \\
\label{eq:sigma0B}
\sigma_0^2&=N_p(N_p-1)T\int df \frac{\Gamma^2 S_{h}^2 P^2}{\left[ (P+S_{h})^2+\Gamma^2S_{h}^2 \right]^2}, \\
\label{eq:sigma1B}
\sigma_1^2&=N_p(N_p-1)T\int df \frac{\Gamma^2 S_{h}^2 \left[ (P+S_{h})^2+\Gamma^2S_{h}^2 \right]}{\left[(P+S_{h})^2+\Gamma^2S_{h}^2 \right]^2},
\end{align}
%%%%%%%%%%%%%%%%%%%%%
%
where we have made the further assumption that the signal's spectrum $S_h$ is known and matched to a template $S_{h0}=S_h$, and we have replaced the pulsar-pair dependent correlation function $\Gamma_{ij}$ 
\citep[also known as Hellings \& Downs function,][]{Hellings1983} with the square root of its variance, i.e., $\Gamma=1/(4\sqrt{3})$. The signal's 
power spectral density (PSD) $S_h$ is related to the characteristic strain derived in Section \ref{sec:GW} via:

%%%%%%%%%%%%%%%%%%%%%%%%
\begin{equation}
\label{eq:sh}
S_h=\frac{h_c^2}{12\pi^2f^3}.
\end{equation}
%%%%%%%%%%%%%%%%%%%%%%%%%
%
For the PSD of the noise $P$, assumed to be the same for all pulsars, we use the form

%%%%%%%%%%%%%%%%%%%%%%%%
\begin{equation}
\label{eqpicorr}
P=2\sigma^2\Delta t+\frac{\delta}{f^5},
\end{equation}
%%%%%%%%%%%%%%%%%%%%%%%
%
where $\sigma=100$ns is the rms residual of the measured TOAs, $\Delta{t}$ is the assumed cadence of individual MSPs observations, and

%%%%%%%%%%%%%%%%%%%%%%%
\begin{equation}
\label{eqdelta}
\delta=5\times10^{-49} \left(\frac{10{\rm yr}}{T}\right)^5\left(\frac{\sigma}{100{\rm ns}}\right)^2\frac{\Delta t}{2\,{\rm weeks}}.
\end{equation}
%%%%%%%%%%%%%%%%%%%%%%
%
With this prescription, the second term on the right-hand side of equation (\ref{eqpicorr}) mimics the loss of sensitivity seen in current PTAs at low frequency and due to fitting the MSP spin 
and spin derivative in the timing model. Note that we do not include any red-noise contribution to the noise PSD, which can be easily accounted for by adding a suitable term $P_{\rm rn}(f)$ in equation (\ref{eqpicorr}). 
For each value of $A$ drawn from the GWB-amplitude distribution reported in \citet{Sesana2016}, we compute the expected $h_c(f)$ for {\it Model-delayed} and {\it Model-stalled} as explained in point (ii) above, 
and for each value of $N_p$ and $T$ we compute the DP using equation (\ref{eq:DP}).

Results are shown in figures \ref{fig:hc_shank} and \ref{fig:dp_shank}. Figure \ref{fig:hc_shank} shows the distribution of the expected amplitudes as a function of frequency. 
In {\it Model-stalled}, the signal is generally flatter than the canonical $f^{-2/3}$ power law in the relevant PTA frequency range, thus 
it is not sufficient to simply report GWB amplitudes $A$ when quoting results. This is also true in {\it Model-delayed}, even though the departure from $f^{-2/3}$ is minimal at $f>1$nHz. 
The {\it Model-stalled} amplitude range still spans more than an order of magnitude and is shifted down by about a factor of two at the currently most relevant PTA frequencies (marked by the stars) 
compared to the fiducial model. The whole predicted range ($99.7\%$ confidence region) is below the current best PTA limit \citep{Shannon2015}, but well within the reach of a putative SKA array under
our conservative assumptions.

This is better quantified in figure \ref{fig:dp_shank}, that shows the DP in the observation time ($T$) -- number of pulsars ($N_p$) plane. For a given $N_p$, 
the $50\%$ DP timescale is delayed by only 3-to-6 years in the {\it Model-stalled} scenario. 
The plot highlights the importance of having a sufficiently large $N_p$, i.e. a larger array helps to narrow this time gap. In fact, detection is based on correlation statistics, 
which is very sensitive to the number of pulsar pairs that can be correlated.
We see that if $N_p=5$, even at $T=20$ yr we still have only DP $\approx 0.3$. A larger array of $N_p=50$, instead, reaches the same DP value after only 10yr and by $T=20$ yr has DP $> 0.95$. Overall, 
these findings support the statement that PTAs {\it will} eventually detect the stochastic GWB from MBHBs, regardless of possible binary stalling issues.

%%%%%%%%%%%%%%%%%%%%%%%%%%%%%%%%%%%%%%%%%%%%%%%%%%%%%%%%%%%%%%%%%%%%%%%%%%%%%%%%%%%%%
%%%%%%%%%%%%%%%%%%%%%%%%%%%%%%%%%%%%%%%%%%%%%%%%%%%%%%%%%%%%%%%%%%%%%%%%%%%%%%%%%%%%%
\section{Caveats}
\label{sec:discussion}

%%%%%%%%%%%%%%%%%%%%%%%%%%%%%%%%%%%%%%%%%
\begin{figure}
\includegraphics[scale=0.41,clip=true]{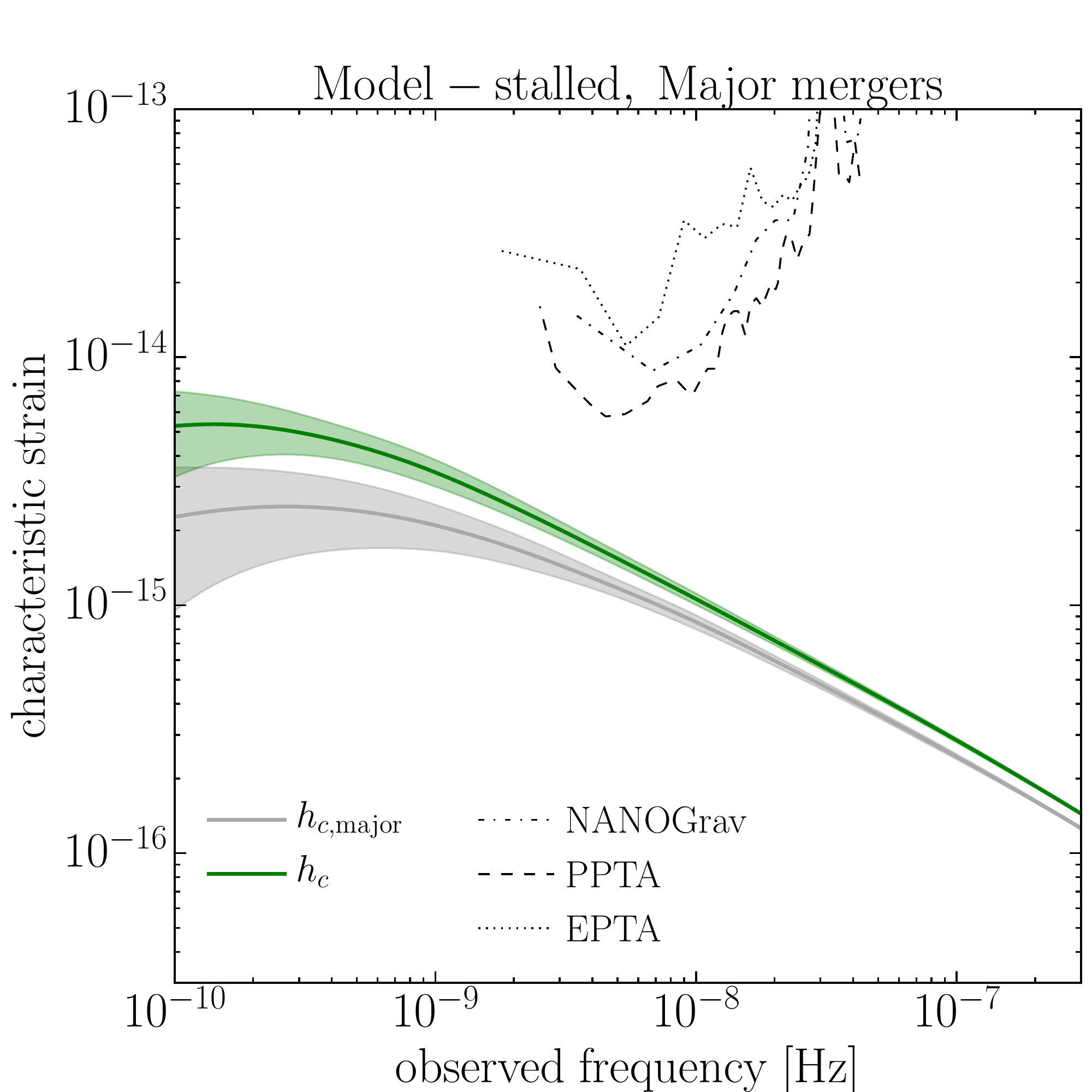}
\caption{GWB spectrum $h_c$ of {\it Model-stalled} when considering only major mergers, i.e., when $q_{\rm in} \geq 0.1$ and $q_{\rm out} \geq 0.1$ (grey line and area) 
compared to the total GWB predicted by the model (green line and area). Lines and shaded areas have the same meaning as in figure \ref{fig:hc2panel}.}
\label{fig:hcmajor}
\end{figure}
%%%%%%%%%%%%%%%%%%%%%%%%%%%%%%%%%%%%%%%%%

Although employing an accurate treatment of the three-body dynamics including an external potential, dynamical friction, stellar hardening and PN equations of motion consistently 
derived from the 3-body PN Hamiltonian (Paper I), the results presented here are subject to a number of caveats that we briefly discuss in the following.

First, we did not attempt to model the interaction of quadruple MBH systems. This is an important point, especially in {\it Model-stalled} where, necessarily, most galaxy mergers contain 
pairs of MBHBs instead of pairs of single MBHs. In fact, in this case mergers are dominated by quadruple systems (cf table \ref{tab2}). As already mentioned, by removing one body and 
thus recovering a triplet, our estimation of the quadruple contribution to the GWB is conservative. We also stress that even if for some (unexpected) reasons quadruple interactions were to lead to no mergers, 
the total GWB would be further suppressed by a factor of two only, and we therefore conclude that our results are robust against this instance.

Second, in the simulations of Paper II we did not account for the later evolution of the MBHs that are ejected in triple systems. As shown by \citet{Hoffman2007}, those MBHs may fall back to the galactic nucleus
on timescales shorter than the Hubble time, thus giving rise to an additional MBH merger in about $10-20\%$ of cases. For this reason, however, this effect is sub-dominant
relative to the main one (the prompt and GW-driven triplet-induced mergers that we account for in this paper), and is likely to depend on the exact modelling of the galactic
potential (and namely its triaxiality, which is unknown). Moreover, as already mentioned, since we neglect this effect our results should be regarded as conservative.

Third, as stressed in section \ref{sec:description}, our semianalytic galaxy-formation model only includes the dynamical friction timescale of galaxy satellites in the potential well of the primary galaxy until the two merge, but does not model the the early migration of MBHs driven by dynamical friction against the gaseous and stellar distribution, in the early epochs following the merger.
We do this on the grounds that this timescale is generally short relative to that of the dynamical friction between the two halos and galaxies, and relative
to the timescales that describe the evolution of MBHBs at separations $\lesssim$ a few pc (stellar hardening, gas-driven migration, triple MBH interactions), at least for the comparable-mass 
MBHBs  that provide the bulk of the PTA signal~\citep[DB17,][]{Dosopoulou2017}. For this reason, the results of this paper are robust against this assumption,
which anyway affects only the {\it Model-delayed} results and not directly the {\it Model-stalled} ones (where the MBH merger timescales are set to values larger than the Hubble time).
Nevertheless, N-body simulations of galaxy pairs find that merging within an Hubble time might be difficult for MBH systems with mass ratios $\lsim 0.1$. 
Therefore this issue, while formally absent in {\it Model-stalled}, may have consequences also for that scenario, because the simulations of Paper II {\it assume} (as initial conditions) 
that MBHs are efficiently brought down to separations comparable to the primary MBH's sphere of influence. However, we have checked that low mass ratio systems do not contribute 
significantly to our results, even in {\it Model-stalled},  by removing all triplets with either $q_{\rm in}$ or $q_{\rm out}$ lower than 0.1 
from the GWB calculation. Results are shown in figure \ref{fig:hcmajor}. It is clear that low $q$ systems do not significantly contribute to the GWB normalisation, being the signal 
at high frequencies only about $10\%$ lower after their removal. It appears, however, that high $q$ triplets tend to produce more eccentric binaries, causing a higher frequency flattening 
and spectral turnover when low $q$ systems are not included in the calculation. The difference in GWB amplitude is however still less than $50\%$ at frequencies of few nHz, 
relevant to PTA experiments.

Last, when applying our results to the \citet{Sesana2016} GWB signal distribution, we are implicitly assuming that the signal correction factor 
 $C(f)$ due to stalling directly applies to MBHB populations that are different from those produced by our semianalytic model. Indeed, even though the two 
 models both assume observational
 selection effects on the scaling relations,
 the intrinsic MBH-galaxy scaling relations of the populations are not necessarily exactly the same \citep{Barausse2017}.
 Although changing ingredients such the employed scaling relation should not change the occurrence of mergers due to triple interactions (our triplet-induced 
merger fractions are fairly independent on the mass scale of the problem, see Paper II), things can be different for MBH evolution models relying on radically 
different merger histories. For example, if galaxy merger rates (which are relatively poorly constrained by observations) are much less frequent, then 
the occurrence of subsequent mergers is much rarer, implying a lower triplet formation rate. This, in turn, will cause a larger suppression of the GWB. We note, 
however, that in {\it Model-stalled} the majority of mergers actually involve quadruple systems. This means that mergers are frequent enough that, in {\it Model-stalled}, the vast majority 
of galaxies hosts a MBHB {\it at any time} along cosmic history. In practice, only a radically different structure formation scenario in which massive galaxies experience on average less than one merger along the cosmic history would result in a larger suppression of the GW signal in {\it Model-stalled} compared to {\it Model-delayed}. In such scenario, in fact, the paucity of galaxy interactions would imply a very low probability for a given galaxy to experience two subsequent mergers, and MBH triplets would not form. We note, however, that both semianalytic and numerical simulations support a rich merger history for massive galaxies \citep[see, e.g.][]{Volonteri2003,DeLucia2007,Oser2010,Kelley2017b}, making the aforementioned scenario extremely unlikely.
In support to this consideration, we note that by employing a completely different framework and MBH evolution model, \citet{Ryu2017} 
find a very similar suppression in the GWB normalisation of about $30\%$.

Although these are substantial caveats, we argue that they are unlikely to strongly influence the results obtained here and our conclusions are thus robust.

%%%%%%%%%%%%%%%%%%%%%%%%%%%%%%%%%%%%%%%%%%%%%%%%%%%%%%%%%%%%%%%%%%%%%%%%%%%%%%%%%%%%%
%%%%%%%%%%%%%%%%%%%%%%%%%%%%%%%%%%%%%%%%%%%%%%%%%%%%%%%%%%%%%%%%%%%%%%%%%%%%%%%%%%%%%
\section{Conclusions}
\label{sec:conclusions}

We have explored the effect of MBH triple interactions on the GW signal produced by a cosmic population of MBHs. 
To this purpose, we have coupled a large library of numerical simulations of triple interactions (Paper I, Paper II) to a semianalytic model for galaxy- and MBH-evolution \citep{Barausse2012}. The numerical simulations solve the 3-body equations 
of motion consistently derived from the 3-body PN Hamiltonian (Paper I) through 2.5PN order, and include
the  effect of the galactic potential and an analytic treatment of dynamical friction and stellar hardening. The library of outcomes is then 
implemented within the semianalytic model, which keeps track of the evolution of individual MBHs and the formation of MBH binaries, triplets and quadruplets 
following galaxy mergers. This framework has allowed us to assess the effect of triple (and quadruple) interactions on the MBHB cosmic merger rate 
and on the expected GWB in the PTA band. 

In particular, we have considered two models for the dynamics of MBHBs. In our fiducial model (labelled {\it Model-delayed}) MBHBs merge on timescales 
of millions-to-billions years, consistently estimated from the properties of the host galaxy. In the PTA band, most mergers occur in gas-poor galaxies and typical 
merger timescales due to stellar-driven hardening (assuming efficient loss cone replenishment) are of the order of Gyrs. This allows the formation of 
several triple MBH systems due to subsequent galaxy mergers, but the formation of `standard' MBHBs is still the dominant coalescence channel. We have then considered
an extreme model where all standard dynamical processes are inefficient at driving MBHBs to sub-pc scales, and MBHBs stall close to their hardening radius ({\it Model-stalled}). The 
rationale behind this model was to investigate the outcome of the most pessimistic scenario from the GW generation standpoint; naively, if all MBHBs stall,
no (or very little, as pointed out by DB17) GW signal is expected in the PTA band. However, mergers can still be triggered by triple (and quadruple) MBH 
interactions following subsequent galaxy mergers, a possibility that was not accounted for in DB17 and which we have explored here in depth.
%for the first time.

Our main results can be summarised as follows:

%%%%%%%%%%%%%%%%%%%
\begin{enumerate}
\item Even if the final-parsec problem is not naturally solved by the interaction of MBHBs with their stellar and gaseous environment (i.e. in {\it Model-stalled}), triple 
interactions can still lead a large number of MBHBs to final coalescence. In the specific galaxy and MBH evolution model explored here, when stalling is assumed, the MBHB 
merger rate is only suppressed by a factor of $\approx4$ (cf table \ref{tab2}) in the mass relevant to PTA observations. Those mergers are the result of triple interactions.
\item The implied GWB background is only suppressed by a factor of about 2-to-3 in the relevant PTA frequency range $1$ nHz $<f<10$ nHz. 
\item Triple MBH interactions naturally produce eccentric binaries. This causes the GWB to be generally flatter than the standard $f^{-2/3}$ power law. However, 
no clear turnover is seen, at least above $\gsim 0.1$ nHz, due to the wide range of eccentricities of the binaries.
\item The most massive MBHBs ${\cal M}>10^{8.5}\msun$, which are the most likely to be individually resolved, can have eccentricities larger than $0.9$ at the relevant PTA 
frequencies. Still, the majority of them tends to be circular or mildly eccentric.
\item When coupling the GWB suppression due to stalling to a pessimistic MBHB population model that predicts a particularly low GWB, 
we still obtain amplitude normalisations at the level $A\gsim 10^{-16}$.
\item The predicted amplitude is well within the reach of SKA. We find that a putative array monitoring 50 pulsars at $100$ ns level has a $90\%$ chance of detection after 15 years of
observation. In general, we find that stalling will delay GWB detection by only about 3-to-6 years depending on the number of pulsars in the array.
\end{enumerate}
%%%%%%%%%%%%%%%%%%%%
%
A particularly relevant result is that signal amplitudes below $A\approx10^{-16}$ are extremely unlikely even in the most pessimistic scenario in which (i) MBHs are intrinsically less 
massive than predicted by standard MBH-host galaxy relations {\it and} (ii) MBHBs stall.

Our main claim is therefore that, because of triple interactions, stalling does not strongly decrease the level of the GWB in the PTA frequency range. The only other way to pose a
threat to future PTA detections is if the opposite of stalling is realised in Nature; i.e. if an extremely efficient coupling with the environment swiftly drives MBHBs through the 
PTA band, which would cause a low-frequency turnover in the GWB. To be dangerous, such turnover should be at frequencies well above $10$ nHz, which for realistic environments is never the 
case, unless {\it all} MBHBs are extremely ($e>0.99$) eccentric. This seems a very unlikely possibility since simulations of MBHBs in stellar environments generally find a {\it range} of
eccentricities $0\lesssim e \lesssim 1$. The most important implication is that with the advent of MeerKAT, FAST and SKA, PTAs {\it will} detect a GW signal from merging MBHBs, provided that those 
instruments bring an-order-of-magnitude improvement over current PTA sensitivities.

Our results are subject to a number of caveats that we have extensively discussed: a very approximate treatment of quadruple interactions, our assumption that MBHBs are driven to separations comparable with their influence radius on timescales shorter than the Hubble time, 
the direct application of our findings to different MBHB populations to derive a lower limit for the expected GWB amplitude. We have argued that none of those caveats 
is critical and that the 
results presented here are robust. Even if everything conspires to produce the lowest possible GWB amplitude, a typical SKA-based PTA will still have a $>90\%$ probability 
of detecting a signal within 15 years 
of data collection, which strengthens the scientific case of this observatory and which is a good reason to look with optimism at the future of  GW astrophysics in the nHz band.

%%%%%%%%%%%%%%%%%%%%%%%%%%%%%%%%%%%%%%%%%%%%%%%%%%%%%%%%%%%%%%%%%%%%%%%%%%%%%%%%%%%%%
%%%%%%%%%%%%%%%%%%%%%%%%%%%%%%%%%%%%%%%%%%%%%%%%%%%%%%%%%%%%%%%%%%%%%%%%%%%%%%%%%%%%%
\section*{Acknowledgements}
MB and FH acknowledge partial financial support from the INFN TEONGRAV specific initiative. MB acknowledges the CINECA award under the ISCRA initiative, for the availability of high performance computing resources and support. This work was supported by the H2020-MSCA-RISE-2015 Grant No. StronGrHEP-690904. AS is supported by a University Research Fellowship of the Royal Society. 
This work has made use of the Horizon Cluster, hosted by the Institut d'Astrophysique de Paris. We thank Stephane Rouberol for running smoothly this cluster for us.

%%%%%%%%%%%%%%%%%%%%%%%%%%%%%%%%%%%%%%%%%%%%%%%%%%%%%%%%%%%%%%%%%%%%%%%%%%%%%%%%%%%%%
%%%%%%%%%%%%%%%%%%%%%%%%%%%%%%%%%%%%%%%%%%%%%%%%%%%%%%%%%%%%%%%%%%%%%%%%%%%%%%%%%%%%%
\bibliographystyle{mn2e}
\bibliography{Bibliography}

\begin{thebibliography}{90}
\expandafter\ifx\csname natexlab\endcsname\relax\def\natexlab#1{#1}\fi

\bibitem[{{Antonini}, {Barausse} \& {Silk}(2015{\natexlab{a}}){Antonini},
  {Barausse}, \& {Silk}}]{Antonini2015}
{Antonini} F., {Barausse} E., {Silk} J., 2015{\natexlab{a}}, \apj, 812, 72

\bibitem[{{Antonini}, {Barausse} \& {Silk}(2015{\natexlab{b}}){Antonini},
  {Barausse}, \& {Silk}}]{Antonini_Barausse2015}
{Antonini} F., {Barausse} E., {Silk} J., 2015{\natexlab{b}}, \apjl, 806, L8

\bibitem[{{Arzoumanian} {et~al}\mbox{.}(2016){Arzoumanian}, {Brazier},
  {Burke-Spolaor}, {Chamberlin}, {Chatterjee}, {Christy}, {Cordes}, {Cornish},
  {Crowter}, {Demorest}, {Deng}, {Dolch}, {Ellis}, {Ferdman}, {Fonseca},
  {Garver-Daniels}, {Gonzalez}, {Jenet}, {Jones}, {Jones}, {Kaspi}, {Koop},
  {Lam}, {Lazio}, {Levin}, {Lommen}, {Lorimer}, {Luo}, {Lynch}, {Madison},
  {McLaughlin}, {McWilliams}, {Mingarelli}, {Nice}, {Palliyaguru}, {Pennucci},
  {Ransom}, {Sampson}, {Sanidas}, {Sesana}, {Siemens}, {Simon}, {Stairs},
  {Stinebring}, {Stovall}, {Swiggum}, {Taylor}, {Vallisneri}, {van Haasteren},
  {Wang}, {Zhu}, \& {NANOGrav Collaboration}}]{Arzoumanian2016}
{Arzoumanian} Z. {et~al.}, 2016, \apj, 821, 13

\bibitem[{{Barausse}(2012)}]{Barausse2012}
{Barausse} E., 2012, \mnras, 423, 2533

\bibitem[{{Barausse} {et~al}\mbox{.}(2017){Barausse}, {Shankar}, {Bernardi},
  {Dubois}, \& {Sheth}}]{Barausse2017}
{Barausse} E., {Shankar} F., {Bernardi} M., {Dubois} Y., {Sheth} R.~K., 2017,
  \mnras, 468, 4782

\bibitem[{{Begelman}, {Blandford} \& {Rees}(1980){Begelman}, {Blandford}, \&
  {Rees}}]{Begelman1980}
{Begelman} M.~C., {Blandford} R.~D., {Rees} M.~J., 1980, \nat, 287, 307

\bibitem[{{Blecha} {et~al}\mbox{.}(2016){Blecha}, {Sijacki}, {Kelley},
  {Torrey}, {Vogelsberger}, {Nelson}, {Springel}, {Snyder}, \&
  {Hernquist}}]{Blecha2016}
{Blecha} L. {et~al.}, 2016, \mnras, 456, 961

\bibitem[{{Bonetti} {et~al}\mbox{.}(2016){Bonetti}, {Haardt}, {Sesana}, \&
  {Barausse}}]{Bonetti2016}
{Bonetti} M., {Haardt} F., {Sesana} A., {Barausse} E., 2016, \mnras, 461, 4419

\bibitem[{{Bonetti} {et~al}\mbox{.}(2017){Bonetti}, {Haardt}, {Sesana}, \&
  {Barausse}}]{Bonetti2017b}
{Bonetti} M., {Haardt} F., {Sesana} A., {Barausse} E., 2017, ArXiv e-prints

\bibitem[{{Boylan-Kolchin}, {Ma} \& {Quataert}(2008){Boylan-Kolchin}, {Ma}, \&
  {Quataert}}]{Boylan-Kolchin2008}
{Boylan-Kolchin} M., {Ma} C.-P., {Quataert} E., 2008, \mnras, 383, 93

\bibitem[{{Bromley} {et~al}\mbox{.}(2006){Bromley}, {Kenyon}, {Geller},
  {Barcikowski}, {Brown}, \& {Kurtz}}]{Bromley2006}
{Bromley} B.~C., {Kenyon} S.~J., {Geller} M.~J., {Barcikowski} E., {Brown}
  W.~R., {Kurtz} M.~J., 2006, \apj, 653, 1194

\bibitem[{{Cattaneo} {et~al}\mbox{.}(2006){Cattaneo}, {Dekel}, {Devriendt},
  {Guiderdoni}, \& {Blaizot}}]{Cattaneo2006}
{Cattaneo} A., {Dekel} A., {Devriendt} J., {Guiderdoni} B., {Blaizot} J., 2006,
  \mnras, 370, 1651

\bibitem[{{Chen}, {Sesana} \& {Del Pozzo}(2017){Chen}, {Sesana}, \& {Del
  Pozzo}}]{Chen2017}
{Chen} S., {Sesana} A., {Del Pozzo} W., 2017, \mnras, 470, 1738

\bibitem[{{Cuadra} {et~al}\mbox{.}(2009){Cuadra}, {Armitage}, {Alexander}, \&
  {Begelman}}]{Cuadra2009}
{Cuadra} J., {Armitage} P.~J., {Alexander} R.~D., {Begelman} M.~C., 2009,
  \mnras, 393, 1423

\bibitem[{{De Lucia} \& {Blaizot}(2007)}]{DeLucia2007}
{De Lucia} G., {Blaizot} J., 2007, \mnras, 375, 2

\bibitem[{{Dekel} \& {Birnboim}(2006)}]{Dekel2006}
{Dekel} A., {Birnboim} Y., 2006, \mnras, 368, 2

\bibitem[{{Dekel} {et~al}\mbox{.}(2009){Dekel}, {Birnboim}, {Engel},
  {Freundlich}, {Goerdt}, {Mumcuoglu}, {Neistein}, {Pichon}, {Teyssier}, \&
  {Zinger}}]{Dekel2009}
{Dekel} A. {et~al.}, 2009, \nat, 457, 451

\bibitem[{{Dekel}, {Sari} \& {Ceverino}(2009){Dekel}, {Sari}, \&
  {Ceverino}}]{Dekel2009b}
{Dekel} A., {Sari} R., {Ceverino} D., 2009, \apj, 703, 785

\bibitem[{{Desvignes} {et~al}\mbox{.}(2016){Desvignes}, {Caballero}, {Lentati},
  {Verbiest}, {Champion}, {Stappers}, {Janssen}, {Lazarus}, {Os{\l}owski},
  {Babak}, {Bassa}, {Brem}, {Burgay}, {Cognard}, {Gair}, {Graikou},
  {Guillemot}, {Hessels}, {Jessner}, {Jordan}, {Karuppusamy}, {Kramer},
  {Lassus}, {Lazaridis}, {Lee}, {Liu}, {Lyne}, {McKee}, {Mingarelli},
  {Perrodin}, {Petiteau}, {Possenti}, {Purver}, {Rosado}, {Sanidas}, {Sesana},
  {Shaifullah}, {Smits}, {Taylor}, {Theureau}, {Tiburzi}, {van Haasteren}, \&
  {Vecchio}}]{Desvignes2016}
{Desvignes} G. {et~al.}, 2016, \mnras, 458, 3341

\bibitem[{{Dosopoulou} \& {Antonini}(2017)}]{Dosopoulou2017}
{Dosopoulou} F., {Antonini} F., 2017, \apj, 840, 31

\bibitem[{{Dotti}, {Sesana} \& {Decarli}(2012){Dotti}, {Sesana}, \&
  {Decarli}}]{Dotti2012}
{Dotti} M., {Sesana} A., {Decarli} R., 2012, Advances in Astronomy, 2012,
  940568

\bibitem[{{Dvorkin} \& {Barausse}(2017)}]{Dvorkin2017}
{Dvorkin} I., {Barausse} E., 2017, ArXiv e-prints

\bibitem[{{Ferrarese} \& {Merritt}(2000)}]{Ferrarese2000}
{Ferrarese} L., {Merritt} D., 2000, \apjl, 539, L9

\bibitem[{{Foster} \& {Backer}(1990)}]{Foster1990}
{Foster} R.~S., {Backer} D.~C., 1990, \apj, 361, 300

\bibitem[{{Gebhardt} {et~al}\mbox{.}(2000){Gebhardt}, {Bender}, {Bower},
  {Dressler}, {Faber}, {Filippenko}, {Green}, {Grillmair}, {Ho}, {Kormendy},
  {Lauer}, {Magorrian}, {Pinkney}, {Richstone}, \& {Tremaine}}]{Gebhardt2000}
{Gebhardt} K. {et~al.}, 2000, \apjl, 539, L13

\bibitem[{{Granato} {et~al}\mbox{.}(2004){Granato}, {De Zotti}, {Silva},
  {Bressan}, \& {Danese}}]{Granato2004}
{Granato} G.~L., {De Zotti} G., {Silva} L., {Bressan} A., {Danese} L., 2004,
  \apj, 600, 580

\bibitem[{{Heger} \& {Woosley}(2002)}]{Heger2002}
{Heger} A., {Woosley} S.~E., 2002, \apj, 567, 532

\bibitem[{{Hellings} \& {Downs}(1983)}]{Hellings1983}
{Hellings} R.~W., {Downs} G.~S., 1983, \apjl, 265, L39

\bibitem[{{Hobbs} {et~al}\mbox{.}(2010){Hobbs}, {Archibald}, {Arzoumanian},
  {Backer}, {Bailes}, {Bhat}, {Burgay}, {Burke-Spolaor}, {Champion}, {Cognard},
  {Coles}, {Cordes}, {Demorest}, {Desvignes}, {Ferdman}, {Finn}, {Freire},
  {Gonzalez}, {Hessels}, {Hotan}, {Janssen}, {Jenet}, {Jessner}, {Jordan},
  {Kaspi}, {Kramer}, {Kondratiev}, {Lazio}, {Lazaridis}, {Lee}, {Levin},
  {Lommen}, {Lorimer}, {Lynch}, {Lyne}, {Manchester}, {McLaughlin}, {Nice},
  {Oslowski}, {Pilia}, {Possenti}, {Purver}, {Ransom}, {Reynolds}, {Sanidas},
  {Sarkissian}, {Sesana}, {Shannon}, {Siemens}, {Stairs}, {Stappers},
  {Stinebring}, {Theureau}, {van Haasteren}, {van Straten}, {Verbiest},
  {Yardley}, \& {You}}]{Hobbs2010}
{Hobbs} G. {et~al.}, 2010, Classical and Quantum Gravity, 27, 084013

\bibitem[{{Hoffman} \& {Loeb}(2007)}]{Hoffman2007}
{Hoffman} L., {Loeb} A., 2007, \mnras, 377, 957

\bibitem[{{Hogg}(1999)}]{Hogg1999}
{Hogg} D.~W., 1999, ArXiv Astrophysics e-prints

\bibitem[{{Iwasawa}, {Funato} \& {Makino}(2006){Iwasawa}, {Funato}, \&
  {Makino}}]{Iwasawa2006}
{Iwasawa} M., {Funato} Y., {Makino} J., 2006, \apj, 651, 1059

\bibitem[{{Jaffe} \& {Backer}(2003)}]{Jaffe2003}
{Jaffe} A.~H., {Backer} D.~C., 2003, \apj, 583, 616

\bibitem[{{Janssen} {et~al}\mbox{.}(2015){Janssen}, {Hobbs}, {McLaughlin},
  {Bassa}, {Deller}, {Kramer}, {Lee}, {Mingarelli}, {Rosado}, {Sanidas},
  {Sesana}, {Shao}, {Stairs}, {Stappers}, \& {Verbiest}}]{Janssen2015}
{Janssen} G. {et~al.}, 2015, Advancing Astrophysics with the Square Kilometre
  Array (AASKA14), 37

\bibitem[{{Jenet} {et~al}\mbox{.}(2006){Jenet}, {Hobbs}, {van Straten},
  {Manchester}, {Bailes}, {Verbiest}, {Edwards}, {Hotan}, {Sarkissian}, \&
  {Ord}}]{Jenet2006}
{Jenet} F.~A. {et~al.}, 2006, \apj, 653, 1571

\bibitem[{{Kauffmann} \& {Haehnelt}(2000)}]{Kauffmann2000}
{Kauffmann} G., {Haehnelt} M., 2000, \mnras, 311, 576

\bibitem[{{Kelley}, {Blecha} \& {Hernquist}(2017){Kelley}, {Blecha}, \&
  {Hernquist}}]{Kelley2017b}
{Kelley} L.~Z., {Blecha} L., {Hernquist} L., 2017, \mnras, 464, 3131

\bibitem[{{Kelley} {et~al}\mbox{.}(2017){Kelley}, {Blecha}, {Hernquist},
  {Sesana}, \& {Taylor}}]{Kelley2017}
{Kelley} L.~Z., {Blecha} L., {Hernquist} L., {Sesana} A., {Taylor} S.~R., 2017,
  \mnras, 471, 4508

\bibitem[{{Khan} {et~al}\mbox{.}(2012){Khan}, {Preto}, {Berczik}, {Berentzen},
  {Just}, \& {Spurzem}}]{Khan2012}
{Khan} F.~M., {Preto} M., {Berczik} P., {Berentzen} I., {Just} A., {Spurzem}
  R., 2012, \apj, 749, 147

\bibitem[{{Klein} {et~al}\mbox{.}(2016){Klein}, {Barausse}, {Sesana},
  {Petiteau}, {Berti}, {Babak}, {Gair}, {Aoudia}, {Hinder}, {Ohme}, \&
  {Wardell}}]{Klein2016}
{Klein} A. {et~al.}, 2016, \prd, 93, 024003

\bibitem[{{Kormendy} \& {Richstone}(1995)}]{Kormendy1995}
{Kormendy} J., {Richstone} D., 1995, \araa, 33, 581

\bibitem[{{Kozai}(1962)}]{Kozai1962}
{Kozai} Y., 1962, \aj, 67, 591

\bibitem[{{Kulier} {et~al}\mbox{.}(2015){Kulier}, {Ostriker}, {Natarajan},
  {Lackner}, \& {Cen}}]{Kulier2015}
{Kulier} A., {Ostriker} J.~P., {Natarajan} P., {Lackner} C.~N., {Cen} R., 2015,
  \apj, 799, 178

\bibitem[{{Kulkarni} \& {Loeb}(2012)}]{Kulkarni2012}
{Kulkarni} G., {Loeb} A., 2012, \mnras, 422, 1306

\bibitem[{{Lapi} {et~al}\mbox{.}(2014){Lapi}, {Raimundo}, {Aversa}, {Cai},
  {Negrello}, {Celotti}, {De Zotti}, \& {Danese}}]{Lapi2014}
{Lapi} A., {Raimundo} S., {Aversa} R., {Cai} Z.-Y., {Negrello} M., {Celotti}
  A., {De Zotti} G., {Danese} L., 2014, \apj, 782, 69

\bibitem[{{Lentati} {et~al}\mbox{.}(2015){Lentati}, {Taylor}, {Mingarelli},
  {Sesana}, {Sanidas}, {Vecchio}, {Caballero}, {Lee}, {van Haasteren}, {Babak},
  {Bassa}, {Brem}, {Burgay}, {Champion}, {Cognard}, {Desvignes}, {Gair},
  {Guillemot}, {Hessels}, {Janssen}, {Karuppusamy}, {Kramer}, {Lassus},
  {Lazarus}, {Liu}, {Os{\l}owski}, {Perrodin}, {Petiteau}, {Possenti},
  {Purver}, {Rosado}, {Smits}, {Stappers}, {Theureau}, {Tiburzi}, \&
  {Verbiest}}]{Lentati2015}
{Lentati} L. {et~al.}, 2015, \mnras, 453, 2576

\bibitem[{{Lidov}(1962)}]{Lidov1962}
{Lidov} M.~L., 1962, \planss, 9, 719

\bibitem[{{MacFadyen} \& {Milosavljevi{\'c}}(2008)}]{MacFadyen2008}
{MacFadyen} A.~I., {Milosavljevi{\'c}} M., 2008, \apj, 672, 83

\bibitem[{{Madau}, {Haardt} \& {Dotti}(2014){Madau}, {Haardt}, \&
  {Dotti}}]{Madau2014}
{Madau} P., {Haardt} F., {Dotti} M., 2014, \apjl, 784, L38

\bibitem[{{Madau} \& {Rees}(2001)}]{Madau2001}
{Madau} P., {Rees} M.~J., 2001, \apjl, 551, L27

\bibitem[{{Magorrian} {et~al}\mbox{.}(1998){Magorrian}, {Tremaine},
  {Richstone}, {Bender}, {Bower}, {Dressler}, {Faber}, {Gebhardt}, {Green},
  {Grillmair}, {Kormendy}, \& {Lauer}}]{Magorrian1998}
{Magorrian} J. {et~al.}, 1998, \aj, 115, 2285

\bibitem[{{McWilliams}, {Ostriker} \& {Pretorius}(2014){McWilliams},
  {Ostriker}, \& {Pretorius}}]{McWilliams2014}
{McWilliams} S.~T., {Ostriker} J.~P., {Pretorius} F., 2014, \apj, 789, 156

\bibitem[{{Nixon} {et~al}\mbox{.}(2011){Nixon}, {Cossins}, {King}, \&
  {Pringle}}]{Nixon2011}
{Nixon} C.~J., {Cossins} P.~J., {King} A.~R., {Pringle} J.~E., 2011, \mnras,
  412, 1591

\bibitem[{{Oser} {et~al}\mbox{.}(2010){Oser}, {Ostriker}, {Naab}, {Johansson},
  \& {Burkert}}]{Oser2010}
{Oser} L., {Ostriker} J.~P., {Naab} T., {Johansson} P.~H., {Burkert} A., 2010,
  \apj, 725, 2312

\bibitem[{{Parkinson}, {Cole} \& {Helly}(2008){Parkinson}, {Cole}, \&
  {Helly}}]{Parkinson2008}
{Parkinson} H., {Cole} S., {Helly} J., 2008, \mnras, 383, 557

\bibitem[{Peters \& Mathews(1963)}]{Peters1963}
Peters P.~C., Mathews J., 1963, Phys. Rev., 131, 435

\bibitem[{{Phinney}(2001)}]{Phinney2001}
{Phinney} E.~S., 2001, ArXiv Astrophysics e-prints

\bibitem[{{Press} \& {Schechter}(1974)}]{Press1974}
{Press} W.~H., {Schechter} P., 1974, \apj, 187, 425

\bibitem[{{Quinlan}(1996)}]{Quinlan1996}
{Quinlan} G.~D., 1996, \na, 1, 35

\bibitem[{{Rajagopal} \& {Romani}(1995)}]{Rajagopal1995}
{Rajagopal} M., {Romani} R.~W., 1995, \apj, 446, 543

\bibitem[{{Ravi} {et~al}\mbox{.}(2012){Ravi}, {Wyithe}, {Hobbs}, {Shannon},
  {Manchester}, {Yardley}, \& {Keith}}]{Ravi2012}
{Ravi} V., {Wyithe} J.~S.~B., {Hobbs} G., {Shannon} R.~M., {Manchester} R.~N.,
  {Yardley} D.~R.~B., {Keith} M.~J., 2012, \apj, 761, 84

\bibitem[{{Ravi} {et~al}\mbox{.}(2015){Ravi}, {Wyithe}, {Shannon}, \&
  {Hobbs}}]{Ravi2015}
{Ravi} V., {Wyithe} J.~S.~B., {Shannon} R.~M., {Hobbs} G., 2015, \mnras, 447,
  2772

\bibitem[{{Reardon} {et~al}\mbox{.}(2016){Reardon}, {Hobbs}, {Coles}, {Levin},
  {Keith}, {Bailes}, {Bhat}, {Burke-Spolaor}, {Dai}, {Kerr}, {Lasky},
  {Manchester}, {Os{\l}owski}, {Ravi}, {Shannon}, {van Straten}, {Toomey},
  {Wang}, {Wen}, {You}, \& {Zhu}}]{Reardon2016}
{Reardon} D.~J. {et~al.}, 2016, \mnras, 455, 1751

\bibitem[{{Rodriguez-Gomez} {et~al}\mbox{.}(2015){Rodriguez-Gomez}, {Genel},
  {Vogelsberger}, {Sijacki}, {Pillepich}, {Sales}, {Torrey}, {Snyder},
  {Nelson}, {Springel}, {Ma}, \& {Hernquist}}]{Rodriguez-Gomez2015}
{Rodriguez-Gomez} V. {et~al.}, 2015, \mnras, 449, 49

\bibitem[{{Rosado}, {Sesana} \& {Gair}(2015){Rosado}, {Sesana}, \&
  {Gair}}]{Rosado2015}
{Rosado} P.~A., {Sesana} A., {Gair} J., 2015, \mnras, 451, 2417

\bibitem[{{Ryu} {et~al}\mbox{.}(2017){Ryu}, {Perna}, {Haiman}, {Ostriker}, \&
  {Stone}}]{Ryu2017}
{Ryu} T., {Perna} R., {Haiman} Z., {Ostriker} J.~P., {Stone} N.~C., 2017, ArXiv
  e-prints

\bibitem[{{Sazhin}(1978)}]{Sazhin1978}
{Sazhin} M.~V., 1978, \sovast, 22, 36

\bibitem[{{Sesana}(2013)}]{Sesana2013}
{Sesana} A., 2013, \mnras, 433, L1

\bibitem[{{Sesana}(2015)}]{Sesana2015b}
{Sesana} A., 2015, in Astrophysics and Space Science Proceedings, Vol.~40,
  Gravitational Wave Astrophysics, {Sopuerta} C.~F., ed., p. 147

\bibitem[{{Sesana} {et~al}\mbox{.}(2014){Sesana}, {Barausse}, {Dotti}, \&
  {Rossi}}]{Sesana2014}
{Sesana} A., {Barausse} E., {Dotti} M., {Rossi} E.~M., 2014, \apj, 794, 104

\bibitem[{{Sesana}, {Haardt} \& {Madau}(2006){Sesana}, {Haardt}, \&
  {Madau}}]{Sesana2006}
{Sesana} A., {Haardt} F., {Madau} P., 2006, \apj, 651, 392

\bibitem[{{Sesana} {et~al}\mbox{.}(2004){Sesana}, {Haardt}, {Madau}, \&
  {Volonteri}}]{Sesana2004}
{Sesana} A., {Haardt} F., {Madau} P., {Volonteri} M., 2004, \apj, 611, 623

\bibitem[{{Sesana} \& {Khan}(2015)}]{Sesana2015}
{Sesana} A., {Khan} F.~M., 2015, \mnras, 454, L66

\bibitem[{{Sesana} {et~al}\mbox{.}(2016){Sesana}, {Shankar}, {Bernardi}, \&
  {Sheth}}]{Sesana2016}
{Sesana} A., {Shankar} F., {Bernardi} M., {Sheth} R.~K., 2016, \mnras, 463, L6

\bibitem[{{Sesana}, {Vecchio} \& {Colacino}(2008){Sesana}, {Vecchio}, \&
  {Colacino}}]{Sesana_Vecchio2008}
{Sesana} A., {Vecchio} A., {Colacino} C.~N., 2008, \mnras, 390, 192

\bibitem[{{Sesana}, {Vecchio} \& {Volonteri}(2009){Sesana}, {Vecchio}, \&
  {Volonteri}}]{Sesana2009}
{Sesana} A., {Vecchio} A., {Volonteri} M., 2009, \mnras, 394, 2255

\bibitem[{{Shankar} {et~al}\mbox{.}(2016){Shankar}, {Bernardi}, {Sheth},
  {Ferrarese}, {Graham}, {Savorgnan}, {Allevato}, {Marconi}, {L{\"a}sker}, \&
  {Lapi}}]{Shankar2016}
{Shankar} F. {et~al.}, 2016, \mnras, 460, 3119

\bibitem[{{Shannon} {et~al}\mbox{.}(2015){Shannon}, {Ravi}, {Lentati}, {Lasky},
  {Hobbs}, {Kerr}, {Manchester}, {Coles}, {Levin}, {Bailes}, {Bhat},
  {Burke-Spolaor}, {Dai}, {Keith}, {Os{\l}owski}, {Reardon}, {van Straten},
  {Toomey}, {Wang}, {Wen}, {Wyithe}, \& {Zhu}}]{Shannon2015}
{Shannon} R.~M. {et~al.}, 2015, Science, 349, 1522

\bibitem[{{Siemens} {et~al}\mbox{.}(2013){Siemens}, {Ellis}, {Jenet}, \&
  {Romano}}]{Siemens2013}
{Siemens} X., {Ellis} J., {Jenet} F., {Romano} J.~D., 2013, Classical and
  Quantum Gravity, 30, 224015

\bibitem[{{Taffoni} {et~al}\mbox{.}(2003){Taffoni}, {Mayer}, {Colpi}, \&
  {Governato}}]{Taffoni2003}
{Taffoni} G., {Mayer} L., {Colpi} M., {Governato} F., 2003, \mnras, 341, 434

\bibitem[{{Taylor}(1992)}]{Taylor1992}
{Taylor} J.~H., 1992, Philosophical Transactions of the Royal Society of London
  Series A, 341, 117

\bibitem[{{Taylor} {et~al}\mbox{.}(2016{\natexlab{a}}){Taylor}, {Huerta},
  {Gair}, \& {McWilliams}}]{Taylor2016}
{Taylor} S.~R., {Huerta} E.~A., {Gair} J.~R., {McWilliams} S.~T.,
  2016{\natexlab{a}}, \apj, 817, 70

\bibitem[{{Taylor}, {Simon} \& {Sampson}(2017){Taylor}, {Simon}, \&
  {Sampson}}]{Taylor2017}
{Taylor} S.~R., {Simon} J., {Sampson} L., 2017, Physical Review Letters, 118,
  181102

\bibitem[{{Taylor} {et~al}\mbox{.}(2016{\natexlab{b}}){Taylor}, {Vallisneri},
  {Ellis}, {Mingarelli}, {Lazio}, \& {van Haasteren}}]{Taylor2016b}
{Taylor} S.~R., {Vallisneri} M., {Ellis} J.~A., {Mingarelli} C.~M.~F., {Lazio}
  T.~J.~W., {van Haasteren} R., 2016{\natexlab{b}}, \apjl, 819, L6

\bibitem[{{The NANOGrav Collaboration} {et~al}\mbox{.}(2015){The NANOGrav
  Collaboration}, {Arzoumanian}, {Brazier}, {Burke-Spolaor}, {Chamberlin},
  {Chatterjee}, {Christy}, {Cordes}, {Cornish}, {Crowter}, {Demorest}, {Dolch},
  {Ellis}, {Ferdman}, {Fonseca}, {Garver-Daniels}, {Gonzalez}, {Jenet},
  {Jones}, {Jones}, {Kaspi}, {Koop}, {Lam}, {Lazio}, {Levin}, {Lommen},
  {Lorimer}, {Luo}, {Lynch}, {Madison}, {McLaughlin}, {McWilliams}, {Nice},
  {Palliyaguru}, {Pennucci}, {Ransom}, {Siemens}, {Stairs}, {Stinebring},
  {Stovall}, {Swiggum}, {Vallisneri}, {van Haasteren}, {Wang}, \&
  {Zhu}}]{NANOGrav2015}
{The NANOGrav Collaboration} {et~al.}, 2015, \apj, 813, 65

\bibitem[{{Vasiliev}, {Antonini} \& {Merritt}(2015){Vasiliev}, {Antonini}, \&
  {Merritt}}]{Vasiliev2015}
{Vasiliev} E., {Antonini} F., {Merritt} D., 2015, \apj, 810, 49

\bibitem[{{Verbiest} {et~al}\mbox{.}(2016){Verbiest}, {Lentati}, {Hobbs}, {van
  Haasteren}, {Demorest}, {Janssen}, {Wang}, {Desvignes}, {Caballero}, {Keith},
  {Champion}, {Arzoumanian}, {Babak}, {Bassa}, {Bhat}, {Brazier}, {Brem},
  {Burgay}, {Burke-Spolaor}, {Chamberlin}, {Chatterjee}, {Christy}, {Cognard},
  {Cordes}, {Dai}, {Dolch}, {Ellis}, {Ferdman}, {Fonseca}, {Gair},
  {Garver-Daniels}, {Gentile}, {Gonzalez}, {Graikou}, {Guillemot}, {Hessels},
  {Jones}, {Karuppusamy}, {Kerr}, {Kramer}, {Lam}, {Lasky}, {Lassus},
  {Lazarus}, {Lazio}, {Lee}, {Levin}, {Liu}, {Lynch}, {Lyne}, {Mckee},
  {McLaughlin}, {McWilliams}, {Madison}, {Manchester}, {Mingarelli}, {Nice},
  {Os{\l}owski}, {Palliyaguru}, {Pennucci}, {Perera}, {Perrodin}, {Possenti},
  {Petiteau}, {Ransom}, {Reardon}, {Rosado}, {Sanidas}, {Sesana}, {Shaifullah},
  {Shannon}, {Siemens}, {Simon}, {Smits}, {Spiewak}, {Stairs}, {Stappers},
  {Stinebring}, {Stovall}, {Swiggum}, {Taylor}, {Theureau}, {Tiburzi},
  {Toomey}, {Vallisneri}, {van Straten}, {Vecchio}, {Wang}, {Wen}, {You},
  {Zhu}, \& {Zhu}}]{Verbiest2016}
{Verbiest} J.~P.~W. {et~al.}, 2016, \mnras, 458, 1267

\bibitem[{{Volonteri}, {Haardt} \& {Madau}(2003){Volonteri}, {Haardt}, \&
  {Madau}}]{Volonteri2003}
{Volonteri} M., {Haardt} F., {Madau} P., 2003, \apj, 582, 559

\bibitem[{{Volonteri}, {Lodato} \& {Natarajan}(2008){Volonteri}, {Lodato}, \&
  {Natarajan}}]{Volonteri2008}
{Volonteri} M., {Lodato} G., {Natarajan} P., 2008, \mnras, 383, 1079

\bibitem[{{Wyithe} \& {Loeb}(2003)}]{Wyithe2003}
{Wyithe} J.~S.~B., {Loeb} A., 2003, \apj, 590, 691

\end{thebibliography}
%%%%%%%%%%%%%%%%%%%%%%%%%%%%%%%%%%%%%%%%%%%%%%%%%%%%%%%%%%%%%%%%%%%%%%%%%%%%%%%%%%%%%
%%%%%%%%%%%%%%%%%%%%%%%%%%%%%%%%%%%%%%%%%%%%%%%%%%%%%%%%%%%%%%%%%%%%%%%%%%%%%%%%%%%%%

\end{document}